\documentclass[11pt,a4paper]{article}
\usepackage{xspace}
\usepackage{jheppub}
\usepackage{dsfont}
\usepackage{amsfonts}
\usepackage{amsmath}
\usepackage{amssymb}
\usepackage{bbold}
\usepackage{graphicx}
\usepackage{xurl}
\usepackage{hyperref}
\usepackage{axodraw2}
\usepackage{pstricks}
\usepackage{relsize,exscale,scalefnt,anyfontsize,cases}
\usepackage[utf8]{inputenc}

\def\bc{\begin{center}}
\def\ec{\end{center}}
\def\bi{\begin{itemize}}
\def\ei{\end{itemize}}
\def\be{\begin{equation}}
\def\ee{\end{equation}}
\def\bea{\begin{eqnarray}}
\def\eea{\end{eqnarray}}
\newcommand{\nn}{\nonumber} 
\newcommand{\ie}{{i.e.}}  
\newcommand{\eg}{{e.g.}}

\def\xtwo{x_{_2}}
\newcommand{\unit}{1\!\!1}
\newcommand{\Nc}{\ensuremath{N_c}\xspace}
\newcommand{\morder}[1]{{\cal O}\left(#1 \right)}
\newcommand{\eq}[1]{(\ref{#1})}
\newcommand{\com}[2]{\left[{#1},{#2}\right]}
\newcommand{\ave}[1]{\langle{#1}\rangle}

\newcommand{\tf}{t_{\mathrm{f}}}
\newcommand{\R}{\textnormal{R}}
\newcommand{\Kvec}{{\boldsymbol K}}
\newcommand{\qvec}{{\boldsymbol q}}

\newcommand{\jpsi}{J/\psi}
\newcommand{\pt}{p_{_\perp}}
\newcommand{\pp}{\ensuremath{\text{pp}}\xspace}
\newcommand{\pA}{\ensuremath{\text{pA}}\xspace}
\newcommand{\pPb}{\ensuremath{\text{pPb}}\xspace}
\newcommand{\pN}{p--N}
\newcommand{\dd}{{\rm d}}
\newcommand{\lsim}{\lesssim} 
\newcommand{\gsim}{\gtrsim}

\def\bm#1{\mbox{\boldmath$#1$}}
\newcommand{\tr}{\mathrm{Tr}\,}

\newcommand{\F}{{\cal F}}
\newcommand{\Phat}{\hat{{\cal P}}}
\def\qhat{\hat{q}}

\def \GenericGtoGG  (#1,#2,#3) {
\resizebox{#1 mm}{!}{\raisebox{#2 pt}{
\begin{axopicture}(451,116) (31,-89)
    \SetWidth{1.0}
    \SetColor{Black}
    \Gluon(128.772,-21.27)(45.99,-21.27){2.587}{10}
    \SetWidth{1.0}
    \Line(217.877,-70.134)(223.625,-64.386)\Line(217.877,-64.386)(223.625,-70.134)
    \Line(236.273,-70.134)(242.021,-64.386)\Line(236.273,-64.386)(242.021,-70.134)
    \Gluon(220.751,-67.26)(220.176,-13.222){2.587}{9}
    \Text(462.773,-60)[l]{\fontsize{#3}{2}\selectfont {$p_{\perp} = z K_\perp$}}
    \Text(84.506,-6.324)[l]{\fontsize{#3}{2}\selectfont {$E$}}
    \Gluon(257.543,-67.26)(256.968,-8.623){2.587}{9}
    \Line(254.669,-70.134)(260.417,-64.386)\Line(254.669,-64.386)(260.417,-70.134)
    \SetWidth{1.0}
    \Gluon(441.503,15.522)(128.772,-21.27){2.587}{39}
    \Text(320,-60)[l]{\fontsize{#3}{2}\selectfont {$\xi, \ \Kvec_1 \equiv  \Kvec$}}
    \SetWidth{1.0}
    \Line(48.864,-27.594)(62.661,-31.043)
    \Line(48.289,-19.546)(62.661,-14.372)
    \Line(47.714,-16.096)(60.937,-6.898)
    \Line(32.193,-24.145)(47.714,-24.145)
    \Line(32.193,-18.971)(47.714,-18.971)
    \SetWidth{1.0}
    \GOval(47.714,-22.42)(10.923,5.174)(0){0.882}
    \Gluon(239.147,-67.26)(239.147,-34.492){2.587}{5}
    \Text(320,25)[l]{\fontsize{#3}{2}\selectfont {$1-\xi, \ \Kvec_2 \simeq  -\Kvec$}}
    \Text(459.899,-43.69)[l]{\fontsize{#3}{2}\selectfont {$E_h = z \xi E $}}
    \Oval(239.722,-33.343)(55.188,55.188)(0)
    \SetWidth{1.0}
    \Gluon(413.909,-48.864)(128.772,-21.27){2.587}{36}
    \SetWidth{1.0}
    \Line(450.701,-54.038)(463.923,-54.613)
    \Line(421.382,-53.463)(448.401,-55.188)
    \Line(420.232,-49.439)(450.701,-51.739)
    \Line(420.232,-48.864)(436.904,-44.265)
    \Line(418.508,-45.415)(435.754,-37.367)
    \SetWidth{1.0}
    \GOval(416.783,-50.589)(8.048,7.473)(0){0.882}
    \GOval(449.551,-53.463)(4.599,4.599)(0){0.882}
  \end{axopicture}
}}}

\def \GBsgg  (#1,#2) {
\resizebox{#1 mm}{!}{\raisebox{#2 pt}{  
\begin{axopicture}(228,113) (222,-131)
    \SetWidth{3.0}
    \SetColor{Black}
    \Gluon(312,-43)(256,-127){11}{4}
    \Gluon(416,-127)(366,-42){11}{4}
    \Gluon(448,-31)(224,-31){11}{8}
  \end{axopicture}
  }}}

\def \GBugg  (#1,#2) {
\resizebox{#1 mm}{!}{\raisebox{#2 pt}{  
   \begin{picture}(228,118) (222,-131)
    \SetWidth{3.0}
    \SetColor{Black}
    \Gluon(397,-37)(256,-122){11}{7}
    \Gluon(416,-122)(278,-37){11}{7}
    \Gluon(448,-26)(224,-26){11}{9}
  \end{picture}
  }}}
  
\def \GBtgg  (#1,#2) {
\resizebox{#1 mm}{!}{\raisebox{#2 pt}{    
   \begin{picture}(228,118) (222,-131)
    \SetWidth{3.0}
    \SetColor{Black}
    \Gluon(328,-77)(256,-122){11}{3}
    \Gluon(416,-122)(288,-37){11}{6}
    \Gluon(448,-26)(224,-26){11}{9}
  \end{picture}
  }}}

\def \GBsqg  (#1,#2) {
\resizebox{#1 mm}{!}{\raisebox{#2 pt}{
 \begin{axopicture}(228,111) (222,-133)
    \SetWidth{3.0}
    \SetColor{Black}
    \Line[arrow,arrowpos=0.3,arrowlength=20,arrowwidth=10,arrowinset=0.2](224,-33)(368,-33)
    \Gluon(304,-33)(256,-129){11}{4}
    \Gluon(416,-129)(368,-33){11}{4}
    \Line[arrow,arrowpos=0.75,arrowlength=20,arrowwidth=10,arrowinset=0.2](304,-33)(448,-33)
  \end{axopicture}
  }}}
  
\def \GBuqg  (#1,#2) {
\resizebox{#1 mm}{!}{\raisebox{#2 pt}{
   \begin{axopicture}(228,116) (222,-132)
    \SetWidth{3.0}
    \SetColor{Black}
    \Line[arrow,arrowpos=0.2,arrowlength=20,arrowwidth=10,arrowinset=0.2](224,-28)(368,-28)
    \Gluon(400,-28)(256,-124){11}{6}
    \Gluon(416,-124)(272,-28){11}{6}
    \Line[arrow,arrowpos=0.8,arrowlength=20,arrowwidth=10,arrowinset=0.2](304,-28)(448,-28)
  \end{axopicture}
  }}}

\def \QGstraightidentity  (#1,#2) {
\resizebox{#1 mm}{!}{\raisebox{#2 pt}{  
   \begin{axopicture}(228,89) (222,-148)
    \SetWidth{3.0}
    \SetColor{Black}
    \Line[arrow,arrowpos=0.5,arrowlength=20,arrowwidth=10,arrowinset=0.2](224,-71)(448,-71)
    \Gluon(448,-135)(224,-135){11}{9}
  \end{axopicture}
  }}}

\def \GBsqqbar  (#1,#2) {
\resizebox{#1 mm}{!}{\raisebox{#2 pt}{
  \begin{axopicture}(228,113) (222,-131)
    \SetWidth{3.0}
    \SetColor{Black}
    \Line[arrow,arrowpos=0.7,arrowlength=20,arrowwidth=10,arrowinset=0.2](368,-31)(416,-127)
    \Gluon(310,-43)(256,-127){11}{4}
    \Gluon(368,-31)(224,-31){11}{5}
    \Line[arrow,arrowpos=0.5,arrowlength=20,arrowwidth=10,arrowinset=0.2,flip](368,-31)(448,-31)
  \end{axopicture}
}}}

\def \GBuqqbar  (#1,#2) {
\resizebox{#1 mm}{!}{\raisebox{#2 pt}{
 \begin{axopicture}(228,116) (222,-131)
    \SetWidth{3.0}
    \SetColor{Black}
    \Line[arrow,arrowpos=0.75,arrowlength=20,arrowwidth=10,arrowinset=0.2,flip](304,-28)(448,-28)
    \Gluon(384,-28)(256,-124){11}{6}
    \Gluon(304,-28)(224,-28){11}{3}
    \Line[arrow,arrowpos=0.3,arrowlength=20,arrowwidth=10,arrowinset=0.2,flip](416,-124)(304,-28)
  \end{axopicture}
  }}}

\def \FierzIdentity (#1,#2) {
\resizebox{#1 mm}{!}{\raisebox{#2 pt}{
\fcolorbox{white}{white}{
  \begin{axopicture}(66,60) (223,-202)
    \SetWidth{2.0}
    \SetColor{Black}
    \Line (224,-148)(240,-148)
    \Line (224,-196)(240,-196)
    \Line (288,-148)(272,-148)
    \Line (288,-196)(272,-196)
    \Line[arrow,arrowpos=0.5,arrowlength=12.5,arrowwidth=5,arrowinset=0.2,flip](240,-148)(240,-196)
    \Line[arrow,arrowpos=0.5,arrowlength=12.5,arrowwidth=5,arrowinset=0.2](272,-148)(272,-196)
  \end{axopicture}
}}}}

\def \FierzSinglet (#1,#2) {
\resizebox{#1 mm}{!}{\raisebox{#2 pt}{
\fcolorbox{white}{white}{
  \begin{axopicture}(66,60) (223,-202)
    \SetWidth{2.0}
    \SetColor{Black}
    \Line[arrow,arrowpos=0.5,arrowlength=12.5,arrowwidth=5,arrowinset=0.2,flip](224,-148)(288,-148)
    \Line[arrow,arrowpos=0.5,arrowlength=12.5,arrowwidth=5,arrowinset=0.2](224,-196)(288,-196)
  \end{axopicture}
}}}}

\title{Fully coherent energy loss effects on light hadron production in pA collisions}

\author[a]{Fran\c{c}ois Arleo,}
\author[b]{Florian Cougoulic,}
\author[c]{St\'ephane Peign\'e}

\affiliation[a]{Laboratoire Leprince-Ringuet, \'Ecole polytechnique, Institut polytechnique de Paris, CNRS/IN2P3, 91128 Palaiseau, France}
\affiliation[b]{Department of Physics, The Ohio State University, Columbus, OH 43210, USA}
\affiliation[c]{SUBATECH UMR 6457 (IMT Atlantique, Universit\'e de Nantes, IN2P3/CNRS), 4 rue Alfred Kastler, 44307 Nantes, France}

\emailAdd{francois.arleo@cern.ch}
\emailAdd{cougoulic.1@osu.edu}
\emailAdd{peigne@subatech.in2p3.fr}

\abstract{We single out the role of fully coherent induced gluon radiation on light hadron production in pA collisions. The effect has the same general features as for quarkonium production, however with a richer color structure as the induced radiation depends on the global color charge of the partonic subprocess final state. Baseline predictions for light hadron nuclear suppression in pPb collisions at the LHC are provided, taking into account only the effect of fully coherent energy loss, which proves to be of the same order of magnitude as gluon shadowing or saturation. This underlines the need to include fully coherent energy loss in phenomenological studies of hadron production in pA collisions.}

\keywords{perturbative QCD; proton--nucleus collisions; parton energy loss.}

\begin{document} 

\maketitle
\setcounter{footnote}{0}
\renewcommand{\thefootnote}{\arabic{footnote}} 	

\section{Introduction}

The wealth of hadron production data in proton-nucleus (\pA) collisions at collider (RHIC, LHC) energies allows for a detailed study of parton dynamics in cold nuclear matter and of the various nuclear effects expected to occur in \pA when compared to proton-proton (\pp) collisions. Several formalisms are currently used in phenomenological studies of hadron production in high-energy \pA collisions.

In the collinear factorization approach~\cite{Collins:1989gx}, hadron production cross sections in \pA collisions are evaluated assuming leading-twist QCD factorization and using the nuclear parton distribution functions (nPDFs) of the target nucleus. Collinear factorization is best justified for hadron production at large enough $\pt$, where higher-twist contributions can be neglected. The leading-twist nPDFs are obtained from global fits based on DGLAP evolution~\cite{deFlorian:2011fp,Kovarik:2015cma,Eskola:2016oht,AbdulKhalek:2019mzd}. They exhibit gluon {\it shadowing}, namely a depletion at small $\xtwo \lesssim 10^{-2}$ of the gluon PDF in the nucleus with respect to that in a proton (see Ref.~\cite{Armesto:2006ph} for a review). Gluon shadowing leads to a corresponding depletion of hadron production in pA with respect to \pp collisions, either at RHIC (at forward rapidity) or at LHC~\cite{Helenius:2012wd,QuirogaArias:2010wh}. Note, however, that each nPDF set comes as a collection of predictions (depending on the number of parameters used in each global fit) leading to a large theoretical uncertainty on the quantitative role of shadowing, as a direct consequence of the relative lack of small-$\xtwo$ data currently included in the global fit analyses. 

In the saturation formalism (see~\cite{Gelis:2010nm} a for review), hadron production cross sections in \pA collisions depend on the target unintegrated (\ie, $k_{_\perp}$-dependent) gluon distribution (UGD)~\cite{Albacete:2012xq}.\footnote{Depending on the value of $x_{_1}$ at which the projectile is probed, the  partonic content of the projectile is described either by a UGD (at small enough $x_{_1}$), or by the standard PDF of collinear factorization (at large enough $x_1$) in the so-called hybrid formalism~\cite{Dumitru:2005gt,Altinoluk:2011qy}.} This approach is in principle suitable for describing hadron production at small and moderate transverse momentum, $\pt \sim \morder{Q_s}$, with $Q_s$ the saturation scale in the target nucleus. The UGDs are related to the quark and gluon dipole scattering amplitudes. Starting from some initial conditions at $x_{_0} = 10^{-2}$ incorporating the (classical) effect of nuclear $\pt$-broadening,\footnote{This holds independently of the precise choice for the initial conditions, and stresses that the saturation formalism incorporates additional effects when compared to the leading-twist gluon shadowing of collinear factorization. Note that the initial conditions for the dipole scattering amplitudes may be modelled (as in the MV model~\cite{McLerran:1993ni,McLerran:1993ka}) or determined from a global analysis (using a few model assumptions) of electron-proton collisions at HERA~\cite{Albacete:2010sy}.} the dipole amplitudes are determined at lower $x$ from the JIMWLK equations~\cite{JalilianMarian:1997dw,JalilianMarian:1997gr,Weigert:2000gi,Iancu:2001ad,Iancu:2000hn,Ferreiro:2001qy} encoding small-$x$ quantum evolution. Although the saturation formalism allows one in principle to predict the $x$ and $k_{_\perp}$ dependence of UGDs, it is fair to say that the latter still contain some theoretical uncertainty, arising from the choice of initial conditions and from the approximations used to solve the small-$x$ evolution equations. While first calculations predicted a strong hadron suppression at the LHC~\cite{Albacete:2010bs}, later results pointing to lesser suppression proved to be in agreement, within the theoretical and experimental uncertainties, with LHC pPb measurements~\cite{Tribedy:2011aa,Albacete:2012xq,Rezaeian:2012ye,Lappi:2013zma}.

In addition to these two approaches which aim at describing the gluon distributions in large nuclei, various models have included nuclear effects like $\pt$-broadening~\cite{Kopeliovich:2005ym} or {\it initial-state} parton energy loss in the nuclear medium~\cite{Frankfurt:2007rn,Kang:2012kc}, in order to compute hadron production in pA collisions at RHIC and LHC. To our knowledge, however, up to now no approach has addressed the role of {\it fully coherent energy loss} (FCEL) in cold nuclear matter discussed throughout this paper. FCEL is expected in all processes where the underlying partonic process consists in forward scattering (when viewed in the target nucleus rest frame) of an incoming high-energy parton to an outgoing color charge~\cite{Arleo:2010rb,Peigne:2014uha} or colourful system of partons~\cite{Peigne:2014rka}. It arises from the induced radiation of gluons with formation time $\tf$ much larger than the medium length, $\tf \gg L$. In this regime, the average energy loss becomes proportional to the incoming parton energy $E$, $\Delta E_{_{\textnormal{FCEL}}} \propto E$~\cite{Arleo:2010rb}, thus overwhelming the parton energy loss in the Landau--Pomeranchuk--Migdal regime, $\Delta E_{_{\textnormal{LPM}}} \propto L^2$~\cite{Baier:1996sk,Baier:1996kr,Zakharov:1996fv,Zakharov:1997uu}. FCEL is predicted from first principles in various formalisms~\cite{Arleo:2010rb,Armesto:2012qa,Armesto:2013fca,Peigne:2014uha,Peigne:2014rka,Liou:2014rha,Munier:2016oih} (including the saturation formalism~\cite{Liou:2014rha,Munier:2016oih}\footnote{Let us stress that FCEL and saturation are different effects, \eg\ FCEL plays a crucial role not only at collider but also at fixed-target energies, where $\xtwo$ is not small and saturation effects are absent or negligible.}) and has been shown to be a key effect to understand quarkonium suppression in \pA collisions~\cite{Arleo:2012hn,Arleo:2012rs,Arleo:2013zua}. It is thus natural to investigate the effect of FCEL in other processes, such as open heavy flavour, light hadron, or jet production in \pA collisions. In the present study, a fully detailed version of Ref.~\cite{Arleo:2020eia}, we focus on light hadron single inclusive production.

Our study has several motivations:
\bi
\item{} A primary goal is to set a baseline for the quantitative role of FCEL in light hadron nuclear suppression, by taking into account {\it only} this effect. We show that similarly to quarkonium,  light hadron production in \pA collisions is strongly affected by FCEL, yet with novel features that will be underlined. In particular, FCEL depends on the global color charge of the {\it parton pair} produced in the partonic subprocess. This is a richer situation compared to quarkonium production at low $\pt$~\cite{Arleo:2012hn,Arleo:2012rs,Arleo:2013zua}, where only a {\it color octet} heavy-quark pair is produced in the subprocess. As an interesting consequence, the nuclear suppression of single hadron production is sensitive to the color states of a parton pair, and thus to unusual color factors.
\item{} It has been suggested to use present and future data on hadron production ($h^\pm$~\cite{Helenius:2012wd,QuirogaArias:2010wh}, $D$/$B$ mesons~\cite{Kusina:2017gkz,Eskola:2019bgf}, quarkonia~\cite{Kusina:2017gkz}) in pA collisions as a reliable probe of nPDFs (and of saturation~\cite{Albacete:2012xq}), assuming other physical effects to be negligible. Our study shows that the latter assumption should be reconsidered, due to the presence of sizable FCEL effects. In particular, FCEL should be systematically included in nPDF global fit analyses that use hadron production pA data. In pA collisions, electroweak processes where FCEL is absent~\cite{Arleo:2010rb} should be preferred for a direct extraction of nPDFs, as for instance weak boson~\cite{Paukkunen:2010qg} and Drell-Yan~\cite{Arleo:2015qiv} production. It should also be reminded that no FCEL is expected in deep inelastic scattering (except in the limit of resolved photoproduction~\cite{Arleo:2010rb}), making a future electron-ion collider an ideal probe of nPDF and saturation effects only~\cite{Accardi:2012qut,EIC}.
\item{} Isolating the FCEL effect is also interesting because this effect is associated, as we will see, with a quite small theoretical uncertainty. This results from FCEL being a {\it medium-induced} effect (depending on the difference between coherent radiation spectra in \pA and \pp collisions) fully determined within perturbative QCD. 
\ei

The outline of the paper is as follows. In section~\ref{sec:quarkonium} we remind the basics of the FCEL model for quarkonium production, which is generalized to the case of light hadron production in section~\ref{sec:LHP}. Baseline calculations of FCEL effects on light hadron nuclear suppression, assuming $g\to gg$ forward scattering (which should be dominant at mid-rapidity at LHC), are discussed and compared to experimental data in section~\ref{sec:predictions}. Calculations are generalized in section~\ref{app:other-processes} to $q\to qg$ and $g\to q\bar{q}$ scattering processes. We draw conclusions and mention future studies in section~\ref{sec:discussion}.

\section{Model for quarkonium nuclear suppression: a brief reminder}
\label{sec:quarkonium}

The effect of fully coherent energy loss on quarkonium suppression in \pA collisions was studied previously in Refs.~\cite{Arleo:2012hn,Arleo:2012rs,Arleo:2013zua}. (The effect was also extrapolated to heavy-ion collisions in~\cite{Arleo:2014oha}, in order to get a baseline for cold nuclear matter effects in those collisions.) For quarkonium production at moderate $\pt$ compared to the mass $M$ of the heavy $Q \bar{Q}$ pair, $\pt \lsim M$, and assuming the $Q \bar{Q}$ pair to be produced in a color octet state, the partonic subprocess is similar to $g \to g$ forward scattering (with a final `massive gluon') when viewed in the target nucleus rest frame. The expression of the induced coherent radiation spectrum ${\dd I}/{\dd\omega}$ (with $\omega$ the radiated gluon energy) associated to $1 \to 1$ forward scattering~\cite{Arleo:2010rb,Arleo:2012rs,Peigne:2014uha,Munier:2016oih} is recalled in Appendix~\ref{app:quenching}, see \eq{dIR}--\eq{color-rule}.\footnote{The spectrum associated to $g \to g$ scattering is obtained from \eq{dIR} by setting $F_c = \Nc$.} A crucial feature of FCEL is that the energy spectrum $\omega \, {\dd I}/{\dd\omega}$ scales in $\omega/E$, leading to an average energy loss proportional to the energy $E$ of the radiating color charge~\cite{Arleo:2010rb}, and thus to sizable effects in quarkonium and more generally in hadron production in \pA collisions. 
  
In order to account for FCEL, the quarkonium differential production cross sections in \pp and \pA collisions are related by a shift $\varepsilon$ in the quarkonium energy~\cite{Arleo:2012hn,Arleo:2012rs,Arleo:2013zua}, 
\be
\label{energy-shift}
\frac{1}{A}\frac{\dd\sigma_{\pA}^{\psi}}{\dd E}\left(E, \sqrt{s} \right)  = \int_0^{\varepsilon_{\rm max}} \! \! \dd \varepsilon \,{\cal P}(\varepsilon, E) \, \frac{\dd\sigma_{\pp}^{\psi}}{\dd E} \left( E+\varepsilon, \sqrt{s}\right)\ ,
\ee
where ${\cal P}(\varepsilon, E)$ is the energy loss probability distribution or {\it quenching weight}, with  the energy loss $\varepsilon$ and quarkonium energy $E$ defined in the nucleus rest frame. The upper bound on $\varepsilon$ is $\varepsilon_{\rm max} = {\rm min}(E_{\mathrm{p}} -E, E)$,\footnote{The condition $\varepsilon \leq E_{\mathrm{p}} -E$ follows from energy conservation, and $\varepsilon \leq E$ is imposed for consistency with the soft radiation approximation.} where $E_\mathrm{p} \simeq s/(2 m_\mathrm{p})$ is the projectile proton energy in this frame, with $m_\mathrm{p}$ the proton mass and $\sqrt{s}$ the proton--nucleon collision energy. 

The quenching weight is related to the spectrum ${\dd I}/{\dd \omega}$ as~\cite{Arleo:2012rs}
\be
\label{quenching-spectrum}
{\cal P}(\varepsilon, E)  = \frac{\dd I}{\dd\varepsilon} \, \exp \left\{ - \int_{\varepsilon}^{\infty} \dd\omega  \frac{\dd{I}}{\dd\omega} \right\} = \frac{\partial}{\partial \varepsilon} \, \exp \left\{ - \int_{\varepsilon}^{\infty} \dd\omega  \frac{\dd{I}}{\dd\omega} \right\} 
\equiv \frac{1}{E} \, \hat{\cal P} \left(\frac{\varepsilon}{E} \right) \, , 
\ee 
where $\hat{\cal P}$ is a scaling function of $x \equiv \varepsilon/E$ (a direct consequence of the scaling of $\omega \, {\dd I}/{\dd\omega}$ in $\omega/E$). It also depends on the quarkonium transverse mass $M_{_\perp} = (M^2 + \pt^2)^{\frac{1}{2}}$, and on the nuclear transverse momentum broadening $\ell_{_{\perp \rm A}}$ defined by \eq{gluon-broad}--\eq{qhat-x}. The explicit expression of $\hat{\cal P}$ used in quarkonium production is obtained by setting $F_c = \Nc$ in \eq{quenching-dilog}.

In view of our discussion in section~\ref{sec:LHP}, let us rewrite \eq{energy-shift} in two alternative ways. First, due to the scaling of $\hat{\cal P}$ in $x$, the `energy shift' \eq{energy-shift} can be naturally expressed as a {\it rescaling} of the quarkonium energy by introducing the variable 
\be
\label{zprime-def}
z' \equiv \frac{E}{E+\varepsilon} = \frac{1}{1+x} \, . 
\ee
Changing variable from $\varepsilon$ to $z'$ in \eq{energy-shift} we obtain (the dependence on $\sqrt{s}$ being implicit in the following) 
\be
\label{energy-rescaling-psi}
\frac{1}{A}\frac{\dd\sigma_{\pA}^{\psi}(E)}{\dd E} = \int^1_{z'_{\rm min}} \dd z'  \,{\cal F}_{\mathrm{loss}}(z') \, \frac{\dd\sigma_{\pp}^{\psi}(E/z')}{\dd E} \, , 
\ee
where $z'_{\rm min} = {\rm max}(\textstyle{E/E_\mathrm{p}},\textstyle{1/2})$ and the rescaling probability distribution ${\cal  F}_{\mathrm{loss}}(z')$ reads
\be
\label{F-def}
{\cal  F}_{\mathrm{loss}}(z') = \frac{1}{z^{' \, 2}} \, \hat{\cal P} \left(\frac{1-z'}{z'} \right) \, .
\ee
Second, in \eq{energy-shift} we can trade the quarkonium energy $E$ for the quarkonium rapidity
\be
\label{quarkonium-rap}
y \equiv \frac{1}{2} \ln{\frac{E+p^z}{E-p^z}} = \ln{\frac{E+p^z}{M_{_\perp}}} \simeq \ln{\frac{2E}{M_{_\perp}}} \, ,
\ee
leading to
\be
\label{rapidity-shift-jpsi}
\frac{1}{A}\frac{\dd\sigma_{\pA}^{\psi}(y)}{\dd y} = \int_0^{x_{\rm max}}  \! \! \dd{x} \, \, \frac{\Phat (x)}{1+x} \, \frac{\dd\sigma_{\pp}^{\psi}\left(y+\ln{(1+x)} \right)}{\dd y} \, .
\ee

The equations~\eq{energy-rescaling-psi} and \eq{rapidity-shift-jpsi}, corresponding respectively to an energy rescaling ($E \to E/z'$) and a rapidity shift ($y \to y + \delta$), are equivalent ways to implement FCEL, which will be useful when discussing light hadron production in section~\ref{sec:LHP}. The rapidity shift $\delta$ is related to the rescaling variable $z'$ and fractional energy loss $x$ as
\be
\label{delta-def-0}
\delta = \ln{\frac{1}{z'}} = \ln{(1+x)}  \, .
\ee 
Since $\delta$ is invariant under longitudinal boosts,\footnote{\label{foot:deltamax}This is also the case for the bound $x_{\rm max} = {\rm min}(1, \frac{E_{\mathrm{p}}}{E} -1)$ in \eq{rapidity-shift-jpsi}, which can be expressed as a rapidity difference. Using \eq{quarkonium-rap}, we get $\ln{(E_\mathrm{p}/E)} = y_{\mathrm{max}} - y$, with $y_{\mathrm{max}} \equiv \ln(2 E_\mathrm{p}/M_{_\perp})$ the maximal quarkonium rapidity (at fixed $\pt$). Hence, $x_{\rm max} = {\rm min}(1, e^{y_{\rm max}-y}-1)$ in \eq{rapidity-shift-jpsi}.} the form \eq{rapidity-shift-jpsi} derived in the nucleus target rest frame can be directly transposed to the center-of-mass frame of an elementary proton--nucleon collision.   

The master equation \eq{rapidity-shift-jpsi} was applied to quarkonium production in \pA collisions in Refs.~\cite{Arleo:2012hn,Arleo:2012rs,Arleo:2013zua}. In those studies the \pp cross section is taken as a parametrization fitting the \pp data, and the \pA cross section is obtained from \eq{rapidity-shift-jpsi} by implementing the theoretical prediction for the quenching weight $\hat{\cal P}$. In other words, \eq{rapidity-shift-jpsi} predicts the {\it modification} of the \pp cross section, \ie, the nuclear modification factor 
\be
\label{RpA-psi}
R_{\pA}^{\psi}(y,\pt) = \frac{1}{A} \, {\frac{\dd\sigma_{\pA}^{\psi}}{\dd y \, \dd \pt} \biggr/ \frac{\dd\sigma_{\pp}^{\psi}}{\dd y \, \dd \pt}} 
\ee
expected from the sole effect of fully coherent energy loss. 

As shown in~\cite{Arleo:2012hn,Arleo:2012rs,Arleo:2013zua}, FCEL {\it alone} explains the $\jpsi$ nuclear suppression measured at fixed-target collision energies, $\sqrt{s} \lsim 40$~GeV~\cite{Badier:1983dg,Katsanevas:1987pt,Leitch:1999ea,Abt:2008ya,Arnaldi:2010ky}. This is consistent with the fact that nPDF/saturation effects are expected to be mild/absent at such energies. At collider energies the central prediction of \eq{rapidity-shift-jpsi}~\cite{Arleo:2012rs} (with a narrow theoretical uncertainty band~\cite{Andronic:2015wma}) agrees well with the $\jpsi$ suppression measured in dAu collisions at RHIC ($\sqrt{s}=200$~GeV)~\cite{Adare:2010fn,Adare:2012qf} and \pPb\ collisions at the LHC ($\sqrt{s}=5.02$~TeV)~\cite{Abelev:2013yxa,Adam:2015iga,Aaij:2013zxa}. 
Although experimental uncertainties still leave room for shadowing or saturation in $\jpsi$ suppression at collider energies, the results of the pure FCEL scenario tend to favour minimal estimates of those effects (\eg, nPDF sets with a moderate shadowing). Quite generally, for quarkonium but also for the production of {\it any} hadron species, the predictions with energy loss alone could be used to constrain the magnitude of other nuclear effects.

\section{FCEL in light hadron nuclear suppression}
\label{sec:LHP}

The goal of this section is to single out the FCEL effect in light hadron production in \pA collisions. Since the FCEL quenching weight scales in $\varepsilon/E$ (independently of the process where FCEL occurs), the general procedure to implement this effect will be the same as for quarkonium production, using expressions analogous to \eq{energy-rescaling-psi} or equivalently \eq{rapidity-shift-jpsi}. However, in the case of light hadron production some new features emerge, and for clarity we present below the model in all its details. 

\subsection{Setup and assumptions}
\label{sec:setup}

We consider light hadron single inclusive production in \pp and \pA collisions at large enough $\pt$, namely $\pt \gg \ell_{_{\perp \rm A}}$ (with $\ell_{_{\perp \rm A}}$ the nuclear broadening defined by \eq{gluon-broad}--\eq{qhat-x}), within a standard leading-order (LO) picture.

\begin{figure}[t]
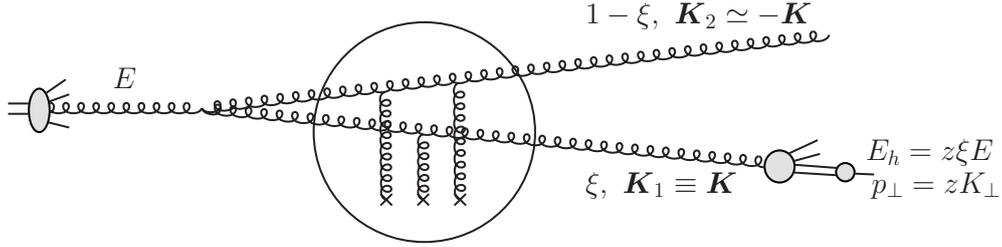

\begin{center}
\hskip -12mm \GenericGtoGG (120,-30,15) 
\end{center}
\caption{Contribution to light hadron production in \pA collisions from the partonic process $g + {\rm A} \to gg + {\rm X}$ followed by gluon fragmentation $g \to h$, as viewed in the target nucleus rest frame.}
\label{fig:typical-process}
\end{figure}

When viewed in the target rest frame, the process arises from $1 \to 2$ forward scattering, where an incoming parton of energy $E$ splits into an outgoing parton pair (also nicknamed `dijet' in the following) being approximately back-to-back in the transverse plane. We will denote the transverse momenta of partons 1 and 2 of the pair as $\Kvec_1 \equiv \Kvec$ and $\Kvec_2 \simeq -\Kvec$, with $K_{_\perp} \equiv |\Kvec| \gg \ell_{_{\perp \rm A}}$, and their energy fractions with respect to the incoming parton as $\xi$ and $1-\xi$. This $1 \to 2$ scattering is followed by the fragmentation of one parton of the pair into the tagged hadron, which thus inherits the transverse momentum $\pt = z K_{_\perp}$, where $z$ is the fragmentation variable, and the energy $E_h= z \xi E$ or $E_h= z (1-\xi) E$ depending on which parton fragments into the hadron. The process is illustrated in Fig.~\ref{fig:typical-process} in the generic case of $g \to gg$ forward scattering, on which we focus in sections \ref{sec:LHP} and \ref{sec:predictions}.  Note that in the c.m.~frame of an elementary proton-nucleon collision, the $g \to gg$ forward scattering is interpreted as the LO $gg \to gg$ partonic subprocess.  

In the target rest frame where energies are very large (transverse momenta being fixed), the gluon energy fractions $\xi$ and $1-\xi$ in the dijet are equivalent to fractions of light-cone $p^+$-momentum (with $p^+ \equiv p^0 + p^z$). The latter fractions are invariant under longitudinal boosts and are related to the rapidity difference of the outgoing gluons, 
\be
\label{y2y1diff}
y_1-y_2  = \log{\left(\frac{\xi}{1-\xi} \right)} \, .
\ee
For further use we also quote the dijet invariant mass:
\be
\label{dijet-mass}
M_{\xi}^2 = 2 K_{_\perp}^2(1+\cosh(y_1-y_2)) = \frac{K_{_\perp}^2}{\xi(1-\xi)} \, .
\ee  

Compared to quarkonium production, light hadron production brings two novelties: (i) the fragmentation variable $z$ between the parent parton and the tagged hadron, and (ii) the possibility to produce the dijet in different {\it color states} (${\rm SU}(\Nc)$ irreducible representations). As we will see in the next section, the former does not bring any complication, whereas the latter makes the FCEL effect richer in light hadron production. Indeed, for a given $1 \to 2$ forward scattering, FCEL depends on the color state of the produced parton pair~\cite{Arleo:2010rb}. This feature requires separating explicitly the different dijet color states in the hadron production cross section, and affecting each of these color states with a different induced energy loss.

\subsection{Implementing fully coherent energy loss}
\label{sec:rescaling}

As recalled in the Introduction, fully coherent radiation is associated with gluon formation times $\tf \gg L$. In general, calculating the induced coherent radiation spectrum associated to the production of a dijet (or multi-parton system) may be complicated. There is however a simplifying limit of this problem, namely, when the induced radiation, in addition to having a formation time $\tf \gg L$, is such that it cannot probe the dijet which thus behaves as a {\it pointlike} color charge. For this `pointlike dijet approximation' (PDA) to hold, two conditions are a priori necessary:
\bi
\item[(i)] At the time $\tf \sim \omega/k_{_\perp}^2$ of its emission, the induced radiation of energy $\omega$ and transverse momentum $k_{_\perp}$ must not probe the transverse size of the parton pair. The pair having a transverse expansion velocity $\sim K_{_\perp}/E$, the latter condition reads
\be
\frac{1}{k_{_\perp}} \gg \frac{\omega}{k_{_\perp}^2} \cdot \frac{K_{_\perp}}{E} \ \Leftrightarrow \ \frac{\omega}{E} \, K_{_\perp} \ll k_{_\perp} \, .
\ee
Since the induced radiation $k_{_\perp}$ must also be softer than the transverse `kick' $\ell_{_{\perp \rm A}}$ suffered by the dijet,\footnote{See the discussion in section 2.2 of Ref.~\cite{Peigne:2014rka}.} we have a fortiori 
\be
\label{condition-1}
\frac{E \, \ell_{_{\perp \rm A}}}{\omega \, K_{_\perp}} \gg 1 \, .
\ee
\item[(ii)] The induced radiation must not probe the color charges of the dijet constituents, but only see the dijet {\it global} color charge. This was shown in Ref.~\cite{Peigne:2014rka} to hold within the logarithmic accuracy\footnote{It was shown in~\cite{Peigne:2014rka} that \eq{log-accuracy} is a {\it sufficient} condition for the induced radiation to depend only on the dijet global color state $\R$. It is not difficult to verify from Ref.~\cite{Peigne:2014rka} that this property is spoiled beyond the logarithmic accuracy \eq{log-accuracy}, so that \eq{log-accuracy} is also a {\it necessary} condition.} 
\be
\label{log-accuracy}
\ln{\left(\frac{E^2 \ell_{_{\perp \rm A}}^2}{\omega^2 K_{_\perp}^2} \right)} \gg 1 \, .
\ee
\ei
The condition \eq{log-accuracy} is stronger than \eq{condition-1} and thus defines the validity domain of the PDA.

The PDA brings major simplifications. First of all, in the PDA the induced spectrum $\omega \, {\dd{I}_{_\R}}/{\dd \omega}$ for a dijet in color state $\R$ is obviously the same as for a pointlike color charge $C_\R$ of mass given by the dijet mass \eq{dijet-mass}. This spectrum and the associated quenching weight $\hat{\cal P}_{_\R}$ are thus obtained from \eq{dIR}--\eq{quenching-dilog} by replacing $M_{_\perp} \to M_{\xi}$. 

Similarly to the case of quarkonium production reviewed in section \ref{sec:quarkonium}, the quenching weight can be traded for the rescaling probability distribution ${\cal  F}_{_\R}(z')$ given by (see \eq{F-def}) 
\be
\label{FR-def}
{\cal  F}_{_\R}(z')= \frac{1}{z^{' \, 2}} \, \hat{\cal P}_{_\R} \left(\frac{1-z'}{z'} \right) \, .
\ee
Since $\F_{_\R}$ depends on $\R$, in the present case the rescaling induced by FCEL must be done for each dijet color state separately. This leads us to express the \pA cross section as an incoherent sum over the accessible dijet color states $\R$, and to introduce the probability $\rho_{_\R}$ for the dijet to be in state $\R$. The $\rho_{_\R}$'s associated to $g \to gg$ are evaluated in Appendix \ref{app:color-proba}, and turn out to depend solely on the energy fraction $\xi$, $\rho_{_{\R}} = \rho_{_{\R}}(\xi)$. Using $1= \sum_{\R}\, \rho_{_{\R}}(\xi)$ we thus write the \pA cross section as
\be
\label{sig-pA-0}
\frac{\dd \sigma_{\rm pA}^{h}(E_h)}{\dd E_h} = \sum_{\R} \, \int \dd{\xi} \, \left[ \rho_{_{\R}}(\xi) \, \frac{\dd \sigma_{\rm pA}^{h}(E_h, \xi)}{\dd E_h\,\dd{\xi}} \right]  \, ,
\ee
where the quantity in brackets is the \pA  cross section to find a hadron of energy $E_h$ in a dijet of color state $\R$ and with energy fractions $\xi$ and $1-\xi$. Note that the latter cross section is inclusive in which parton of the dijet fragments into the tagged hadron. 

Another simplification of the PDA is that the energy rescaling (by a factor $z'$) {\it conserves the dijet internal structure}, and in particular the energy fractions $\xi$ and $1-\xi$. The two partons thus inherit the same rescaling, and in turn the tagged hadron too, independently of the additional rescaling (by a factor $z$) inherent to the fragmentation process.\footnote{This holds because FCEL and fragmentation processes occur at different time scales~\cite{Arleo:2012rs}.} As a consequence, FCEL can be implemented by replacing in \eq{sig-pA-0}, for each color state $\R$, the quantity in brackets by the similar quantity in \pp collisions (multiplied by the atomic mass number $A$), but with $E_h$ rescaled by $z'$ with probability $\F_{_\R}(z')$, 
\be
\label{sig-pA-E} 
\frac{1}{A} \, \frac{\dd \sigma_{\rm pA}^{h}(E_h)}{\dd E_h}  =  \sum_{\R} \, \int \dd{\xi} \, \rho_{_{\R}}(\xi) \int^1_{z'_{\rm min}} \dd{z'} \, \F_{_\R}(z')  \, \frac{\dd \sigma_{\rm pp}^{h}(E_h/z',\xi)}{\dd E_h\,\dd{\xi}}  \, .
\ee
Eq.~\eq{sig-pA-E} generalizes \eq{energy-rescaling-psi} used in quarkonium production to the case of light hadron production. In the PDA the energy rescaling is equivalent to applying the same rapidity shift $\delta$ to the dijet, its constituent partons, and the tagged hadron. Trading the hadron energy $E_h$ for its rapidity $y$ and using \eq{FR-def}, the expression \eq{sig-pA-E} becomes
\be
\label{sig-pA-y} 
\frac{1}{A} \, \frac{\dd \sigma_{\rm pA}^{h}(y,\pt)}{\dd y \, \dd \pt}  =  \sum_{\R} \, \int \dd{\xi} \, \rho_{_{\R}}(\xi) \int_0^{x_{\rm max}} \! \! \frac{\dd{x}}{1+x} \, \, \Phat_{_\R}(x, \ell_{_{\perp \rm A}}, M_{\xi}) \, \, \frac{\dd\sigma_{\pp}^{h}(y+\delta, \pt, \xi)}{\dd y \, \dd \pt \dd{\xi}} \, ,
\ee
where $\delta = \ln{(1+x)}$, $x_{\rm max} = {\rm min}(1, e^{y_{\rm max}-y}-1)$,\footnote{This is the same   as for quarkonium production (see footnote \ref{foot:deltamax}), with $y_{\mathrm{max}}$ being now the maximal hadron rapidity. In the proton--nucleon collision c.m.~frame, $y_{\mathrm{max}} = \ln(\sqrt{s}/\pt)$.} and the dependence of the \pp and \pA cross sections on the hadron $\pt$ is now explicit, as well as that of $\Phat_{_\R}$ on the broadening $\ell_{_{\perp \rm A}}$ and dijet mass $M_{\xi}$. Note that the `hard scale' $K_{_\perp}$ to be used in the expression of $M_{\xi}$ (see \eq{dijet-mass}) reads $K_{_\perp} = \pt/z$, to account for the rescaling of momenta in the fragmentation process. In our approach, the  momentum fraction $z$ of the tagged hadron w.r.t.~its parent parton is treated as a parameter, see section~\ref{sec:procedure}. Eq.~\eq{sig-pA-y} generalizes \eq{rapidity-shift-jpsi} to light hadron production, and will be used in the following to predict the effect of FCEL on the light hadron nuclear modification factor $R_{\pA}$. We have checked that the typical values of $x$ contributing to \eq{sig-pA-y} fulfill the condition \eq{log-accuracy}, thus justifying the PDA and the above implementation of FCEL.  
   
Let us note that the \pp and \pA cross sections appearing in \eq{sig-pA-y} are evaluated at the {\it same} $\pt$. There are two reasons for that. First, as explained in the Introduction, our goal is to single out FCEL, and in \eq{sig-pA-y} we thus neglect the shift in $\pt$ due to nuclear broadening in \pA vs \pp.\footnote{It should be noted that there is no contradiction in isolating the FCEL effect while neglecting $\pt$-broadening, despite the fact that the former is induced by the latter, as illustrated by the vanishing of $\Phat_{_\R}$ when the nuclear broadening $\ell_{_{\perp \rm A}}^2 - \ell_{_{\perp \rm p}}^2$ vanishes, see \eq{quenching-dilog}. In our approach, the role of $\pt$-broadening is simply to specify the quantity  $\ell_{_{\perp \rm A}}$ to be used in the FCEL quenching weight $\Phat_{_\R}$.} Second, the FCEL effect itself can in principle affect the transverse momenta of the dijet constituents, and thus induce a difference between the hadron $\pt$ in the \pA and \pp cross sections. However, this effect should be neglected in the PDA, where in addition to the internal energy fraction $\xi$, the dijet invariant mass \eq{dijet-mass} must be conserved. As a consequence, $K_{_\perp}$ and thus $\pt = z K_{_\perp}$ are conserved. 

To conclude this section, let us stress that our implementation of FCEL consists in a different organization of the perturbative expansion as compared to next-to-leading order (NLO) approaches evaluating absolute \pA cross sections (see \eg\ Refs.~\cite{Chirilli:2011km,Kang:2014lha,Ducloue:2016shw} for single inclusive hadron and Refs.~\cite{Fujii:2013gxa,Ducloue:2015gfa,Ma:2015sia} for quarkonium production). These studies in principle account for the induced radiation of a single gluon, as part of all NLO corrections. In our approach, the induced radiation is resummed to all orders through the expression~\eq{quenching-spectrum} of the quenching weight. This should be meaningful when the spectrum~\eq{dIR} cannot be viewed as a genuine NLO correction, \ie, when $\alpha_s \ln{(E^2 \ell_{_{\perp \rm A}}^2 / \omega^2 K_{_\perp}^2)} \gsim \morder{1}$, consistently with the logarithmic accuracy \eq{log-accuracy} used in our study.

\subsection{Nuclear modification factor $R_{\rm pA}$}
\label{sec:RpA}

The nuclear suppression of a light hadron $h$, in minimum bias \pA collisions as compared to \pp collisions, is commonly represented by the ratio
\be
\label{RpA-1}
R_{\pA}^{h}(y,\pt) = \frac{1}{A} \, {\frac{\dd\sigma_{\pA}^{h}}{\dd y \, \dd \pt} \biggr/ \frac{\dd\sigma_{\pp}^{h}}{\dd y \, \dd \pt}}  \, ,
\ee
where $y$ is the hadron rapidity in the c.m.~frame of an elementary proton--nucleon collision. 

Using \eq{sig-pA-y} we obtain 
\be
\label{RpA-y} 
R_{\pA}^{h}(y,\pt)    =  \sum_{\R} \, \int_0^{x_{\rm max}} \! \!  \frac{\dd{x}}{1+x} \, \, \ave{\rho_{_{\R}}(\xi) \,  \Phat_{_\R}(x, \ell_{_{\perp \rm A}}, M_{\xi})}_{_{y+\delta, \, \pt} } \frac{\ \ \frac{\dd \sigma_{\rm pp}^{h}(y+\delta,\pt)}{\dd y\, \dd \pt}\ \ }{\ \ \frac{\dd \sigma_{\rm pp}^{h}(y,\pt)}{\dd y\, \dd \pt}\ \ } \, ,
\ee
where $\ave{ \  }_{_{y, \, \pt}}$ denotes the $\xi$-average in \pp dijet events where a hadron of rapidity $y$ and transverse momentum $\pt$ is produced, namely,
\be
\label{ave-proba}
\ave{ f(\xi) }_{_{y, \pt}} \equiv  \int \dd{\xi} \, f(\xi) \, g_{_{y, \pt}}(\xi) \ ; \ \ \ g_{_{y, \pt}}(\xi)  =  \frac{\ \ \frac{\dd \sigma_{\rm pp}^{h}(y, \, \pt,\, \xi)}{\dd y\, \dd \pt \dd{\xi}} \ \ }{\int \dd{\xi} \, \frac{\dd\sigma_{\pp}^{h}(y, \, \pt, \, \xi)}{\dd y\, \dd \pt \dd{\xi}}} \, \ \ .
\ee

The functions $\Phat_{_\R}$ and $\rho_{_{\R}}$ being defined in Appendices~\ref{app:quenching} and~\ref{app:color-proba}, the factor $\ave{\rho_{_{\R}} \Phat_{_\R}}$ in \eq{RpA-y} could in principle be evaluated knowing the triple ($y$, $\pt$, $\xi$) differential \pp cross section, \eg\ within collinear factorization once a given set of PDFs and fragmentation functions is chosen. In order to minimize the number of model assumptions, we will instead use the following procedure. 

From the (generalized) mean value theorem, given some $\xi$-average $\ave{ \  }$ defined using a probability density $g(\xi)$, for any continuous function $f(\xi)$ there is a value $\bar{\xi}$ (belonging to the support of $g(\xi)$) such that $\ave{f(\xi)} = f(\bar{\xi})$. The nuclear modification factor~\eq{RpA-y} can thus be written as
\bea
\label{RpA-y-master}
R_{\pA}^{h}(y, \pt, \bar{\xi}) = \sum_{\R} \,  \rho_{_{\R}}(\bar{\xi})  \, R_{\pA}^{\R}(y, \pt, \bar{\xi})  \, ,  \hskip 2.5cm &&  \\ 
\label{RpA-yR}
R_{\pA}^{\R}(y, \pt,\bar{\xi}) = \int_0^{x_{\rm max}} \! \!  \frac{\dd{x}}{1+x} \, \Phat_{_\R}(x, \ell_{_{\perp \rm A}}, M_{\bar{\xi}}) \, \frac{\ \ \frac{\dd \sigma_{\rm pp}^{h}(y+\delta,\pt)}{\dd y\, \dd \pt}\ \ }{\ \ \frac{\dd \sigma_{\rm pp}^{h}(y,\pt)}{\dd y\, \dd \pt}\ \ } \, . \hskip 0.5cm && 
\eea 
Thus, $R_{\pA}^{h}$ is the color average (over the accessible dijet color states $\R$) of the modification factors $R_{\pA}^{\R}$, corresponding to a hadron produced from a dijet in state $\R$. The uncertainty associated to the choice of the parameter $\bar{\xi}$ will be estimated by varying $\bar{\xi}$ in the support interval of $g_{_{y, \pt}}(\xi)$ defined in \eq{ave-proba}. The latter will be chosen as the interval $[0.25,0.75]$, motivated by experimental measurements as discussed in section~\ref{sec:procedure}. 
The above procedure avoids the calculation of $\ave{\rho_{_{\R}} \Phat_{_\R}}$ appearing in \eq{RpA-y} at relatively low cost since the uncertainty band associated to the variation of $\bar{\xi}$ turns out to be quite narrow (see section~\ref{sec:predictions} and Fig.~\ref{fig-uncertainties}). 

We conclude this section by noting that in some special cases of partonic processes, the FCEL spectrum is negative (namely, when $F_c < 0$, see \eq{dIR}--\eq{color-rule}), corresponding to an induced energy {\it gain} in \pA relative to \pp collisions. Then \eq{RpA-yR} should be replaced by
\be
\left. R_{\pA}^{\R}(y, \pt,\bar{\xi}) \right|_{\rm gain}= \int_0^{x_{\rm max}} \! \!  \dd{x} \, (1+x) \, \Phat_{_\R}(x, \ell_{_{\perp \rm A}}, M_{\bar{\xi}}) \, \frac{\ \ \frac{\dd \sigma_{\rm pp}^{h}(y-\delta,\pt)}{\dd y\, \dd \pt}\ \ }{\ \ \frac{\dd \sigma_{\rm pp}^{h}(y,\pt)}{\dd y\, \dd \pt}\ \ } \, ,
\label{RpA-yR-gain}
\ee
where $x_{\rm max} = {\rm min}(1, e^{y_{\rm max}+y}-1)$, and $\Phat_{_\R}$ is associated to the {\it opposite} of the radiation spectrum \eq{dIR} and understood as an energy gain probability density. Eq.~\eq{RpA-yR-gain} is  obtained as follows. In the case of energy gain, the sign of the energy shift (see \eq{energy-shift}) is changed, $E+\varepsilon \to E-\varepsilon$, the rescaling variable $\frac{1}{z'} = \frac{E+\varepsilon}{E} = 1+x$ (see \eq{zprime-def}) becomes $\frac{1}{z'} = 1-x \simeq \frac{1}{1+x}$, and the rapidity shift $\delta = \ln{\frac{1}{z'}} = \ln{(1+x)}$ in \eq{RpA-yR-gain} thus comes with a minus sign.

Eq.~\eq{RpA-yR-gain} applies for instance to the case of the $q \to qg$ forward process (discussed in section \ref{sec:qg}) when the final $qg$ pair is effectively pointlike and color triplet, $\R = {\bf 3}$. Indeed, the color factor \eq{color-rule} then reads $F_c = 2 C_F - \Nc = -1/\Nc <0$. A physical interpretation of fully coherent energy {\it gain} for $q \to q$ scattering can be found in Ref.~\cite{Peigne:2014uha}.

\section{FCEL baseline predictions}
\label{sec:predictions}

We now evaluate the light hadron nuclear suppression due to the sole FCEL effect, based on Eqs.~\eq{RpA-y-master}-\eq{RpA-yR}. In the present section, we assume $g\to gg$ forward scattering to be dominant, which is a reasonable assumption at the LHC, as recalled in section \ref{sec:subprocesses}. The procedure to compute $R_{\pA}^{\R}(y, \pt)$ (for each dijet color state $\R$) and the hadron nuclear modification factor $R_{\pA}^h(y, \pt)$, as well as their associated uncertainties, is presented in section~\ref{sec:procedure}. Results in \pPb collisions at the LHC will be shown in section~\ref{sec:LHC-pheno}, and compared to available experimental data in section \ref{sec:predictions-vs-data}.

\subsection{Partonic subprocess}
\label{sec:subprocesses}

Eqs.~\eqref{RpA-y-master} and \eqref{RpA-yR} have been derived in section \ref{sec:LHP} assuming $g\to gg$ forward scattering, but obviously allow for computing the hadron nuclear modification factor $R_{\pA}^h(y, \pt)$ for any given underlying process (\eg, $q\to qg$, $g\to q\bar{q}$, \ldots). The {\it observable} modification factor could be then obtained by averaging the latter $R_{\pA}^h$'s, with weights given by the relative contributions of each subprocess to the inclusive hadron production cross section. Those weights could in principle be accessed through fixed-order perturbative QCD calculations, but may differ from one calculation to another depending on the choice of factorization scales, parton distributions and fragmentation functions. In practice, however, hadron production at the LHC is dominated by gluon-initiated processes at not too large $\pt$ ($\pt\lesssim50$~GeV) and not too large rapidity ($|y| \lesssim 2$ at $\pt=25$~GeV)~\cite{Sassot:2010bh}, thanks to the huge gluon flux in the projectile and target at small values of $x_{_1}$ and $x_{_2}$. In addition, light hadron production proceeds predominantly by gluon rather than quark fragmentation~\cite{Sassot:2010bh}. 

For these reasons, the hadron nuclear modification factor in \pA collisions at the LHC is determined in the present section assuming the sole $g\to gg$ partonic subprocess, which should be a solid assumption around mid-rapidity. Moreover, we expect the FCEL effect to be qualitatively similar for all partonic subprocesses. In order to illustrate this point, we will give in section~\ref{app:other-processes} our predictions assuming the $q\to qg$ and $g\to q\bar{q}$ channels. The $q\to qg$ channel is interesting in itself because it  becomes important, or even dominant, at $y \gtrsim 3$--$4$, where the projectile hadron is probed at higher momentum fraction, $x_{_1} \propto e^y$, leading to a significant contribution of quark-induced processes. 

It will be important for phenomenology to quantify the FCEL effect in the general case where several partonic processes are of comparable importance in a given rapidity region. This is briefly discussed in section~\ref{sec:discussion} and will be the subject of a future study. 

\subsection{Parameters and theoretical uncertainties}
\label{sec:procedure}

The calculation of $R_{\pA}^{\R}$ follows the  procedure developed for quarkonium studies in Refs.~\cite{Arleo:2012rs,Arleo:2013zua,Arleo:2014oha}. The goal is to determine hadron suppression with the least number of assumptions and parameters, the latter being then varied for a proper determination of theoretical uncertainties.

In order to minimize the model dependence, the double differential \pp cross section $\dd\sigma_{\pp}^{h}/\dd y \, \dd \pt$ entering \eq{RpA-yR} is not taken from theory but determined from a fit to the data. A simple functional form using few parameters allows for an accurate description of the \pp cross section, see Appendix~\ref{app-ppfits}. Since~\eq{RpA-yR} involves a cross section \emph{ratio}, only one of these parameters, namely $n$ in Eq.~\eq{eq:fit}, proves necessary for the calculation of $R_{\pA}^{\R}$. The value used here, $n=15\pm 5$, is determined from a fit to CMS pPb data at $\sqrt{s}=5.02$~TeV~\cite{Khachatryan:2015xaa}. As the only theoretical input in \eq{RpA-yR} is the quenching weight $\Phat_{_\R}$, our calculation of hadron suppression expected from FCEL is directly sensitive to the induced gluon spectrum \eq{dIR}--\eq{color-rule} determined from first principles.
 
The average \eq{RpA-y-master} over the gluon pair (more generally, dijet) color states has a smooth dependence on the parameter $\bar{\xi}$. The default value chosen for $\bar{\xi}$ corresponds to the symmetric configuration of two jets of equal rapidity, leading to $\bar{\xi}=0.5$, which is most likely according to dihadron correlation measurements in \pp collisions at the LHC~\cite{Khachatryan:2016txc}. The uncertainty associated to the choice of  $\bar{\xi}$ is estimated by varying $\bar{\xi}$ by half its default value, $\bar{\xi}=0.50\pm0.25$,\footnote{Note that the dijet invariant mass $M_{\bar{\xi}}$ (entering the quenching weight in \eq{RpA-yR}) is symmetric in $\bar{\xi} \leftrightarrow 1-\bar{\xi}$, and so are the color probabilities $\rho_{_{\R}}$ for the $g\to gg$ channel considered here, leading to the same symmetry for $R_{\pA}^h$, $R_{\pA}^h(y, \pt, \bar{\xi})=R_{\pA}^h(y, \pt, 1-\bar{\xi})$.} which corresponds to a rapidity difference between the two back-to-back jets of approximately $\Delta y=\pm \ln(\bar{\xi}/(1-\bar{\xi}))\simeq \pm1$ unit.
 
The (average) momentum fraction $z$ of the fragmenting parton carried away by the measured hadron depends on the shape of fragmentation functions (and on the hadron species), which are still poorly known. Based on NLO calculations of hadron production at the LHC~\cite{Sassot:2010bh}, we use $z = 0.7 \pm 0.2$  in the calculations to come.

Finally, the central value of the transport coefficient $\hat{q}_{_0}$ is set to $\qhat_{_0} = 0.075$~GeV$^2$/fm (as in Refs.~\cite{Arleo:2012rs,Arleo:2013zua,Arleo:2014oha}), determined from a fit to E866 $\jpsi$ suppression data. The associated uncertainty is determined by varying  $\hat{q}_{_0}$ in the range $0.07$--$0.09$~GeV$^2$/fm~\cite{Arleo:2012rs}.\footnote{In addition to $\hat{q}_{_0}$, the transport coefficient \eq{qhat-x} depends on the momentum fraction $\xtwo$, given by $\xtwo = \frac{\pt}{z \sqrt s} \frac{e^{-y}}{1-\xi_f}$, where $\xi_f$ is defined as the dijet energy fraction carried by the fragmenting parton, either $\xi_f=\xi$ or $\xi_f=1-\xi$. In our procedure, the information about which parton actually fragments into the hadron is not retained (see the comment after \eq{sig-pA-0}). We have checked, however, that the $\xi$ dependence of $\xtwo$ has a marginal effect on $R_{\pA}^{h}$. Discarding this dependence in the following, we set $\xi=0.5$ in the expression of $\xtwo$ and will thus use $\xtwo = \frac{2 \pt}{z\sqrt s} \, e^{-y}$.}

In order to estimate the theoretical uncertainties, these quantities are varied around their central value, $S^0\equiv\{n, \bar{\xi}, z, \hat{q}_{_0}\}$. On top of the central prediction assuming $S^0$, predictions will be made with the sets of parameters $S_k^\pm \equiv  \{p_k^\pm, p_{{i\neq k}}^{}\}$, where the $k^{\rm th}$ parameter is set to its minimal ($p_k^-$) or maximal ($p_k^+$) value, while the other parameters are fixed to their central values.

Assuming that the parameters are uncorrelated, the uncertainty band of our predictions is determined using the Hessian method~\cite{Pumplin:2001ct} (as in~\cite{Arleo:2014oha}) 
 \begin{eqnarray}
\label{eq:errors}
\left(\Delta R_{\pA}^+\right)^2 & = & \sum_k \left[ \max\left\{ R_{\pA}(S^+_k)-R_{\pA}(S^0), R_{\pA}(S^-_k)-R_{\pA}(S^0),0 \right\} \right]^2 \ ,  \nonumber\\
\left(\Delta R_{\pA}^-\right)^2 & = & \sum_k \left[ \max\left\{ R_{\pA}(S^0)-R_{\pA}(S^+_k), R_{\pA}(S^0)-R_{\pA}(S^-_k),0 \right\} \right]^2 \ ,
\end{eqnarray}
where $\{n, n^-, n^+\}=\{15, 10, 20\}$, $\{\bar{\xi},\bar{\xi}^-,\bar{\xi}^+\}=\{0.50,0.25,0.75\}$, $\{z,z^-,z^+\}=\{0.7,0.5,0.9\}$, and (in unit~GeV$^2$/fm) $\{\hat{q}_{_0}, \hat{q}_{_0}^-, \hat{q}_{_0}^+\}=\{0.075, 0.07, 0.09\}$. In the next section we will display (see Fig.~\ref{fig-uncertainties}) the individual contributions $R_{\pA}(S_k^\pm)$ to the total theoretical uncertainty defined in~\eq{eq:errors}.

Finally, as in Ref.~\cite{Arleo:2012rs} we will use $\alpha_s=0.5$ for the strong coupling constant,\footnote{The value of $\alpha_s$ is frozen at the semi-hard scale $\qhat L \lesssim 1$~GeV$^2$ for cold nuclear matter at LHC energies.} $L_{_\textnormal{Pb}}=10.11$~fm for the average path length in the lead nucleus (determined using realistic nuclear density profiles), and $L_{_\textnormal{p}}=1.5$~fm for that in a proton.

\subsection{Results in pPb collisions at the LHC}
\label{sec:LHC-pheno}

Calculations of light hadron suppression in pPb collisions at current top LHC energy, $\sqrt{s}=8.16$~TeV, are shown here as a function of transverse momentum and rapidity. Although somewhat academic, it is instructive to first discuss the nuclear production ratio for a final gluon pair in a given color state ($R_{\pA}^{\R}$ given by \eq{RpA-yR}), and then obtain the `inclusive' hadron suppression through the average \eq{RpA-y-master} over color states.

\begin{figure}[t]
\centering
\includegraphics[width=7.5cm]{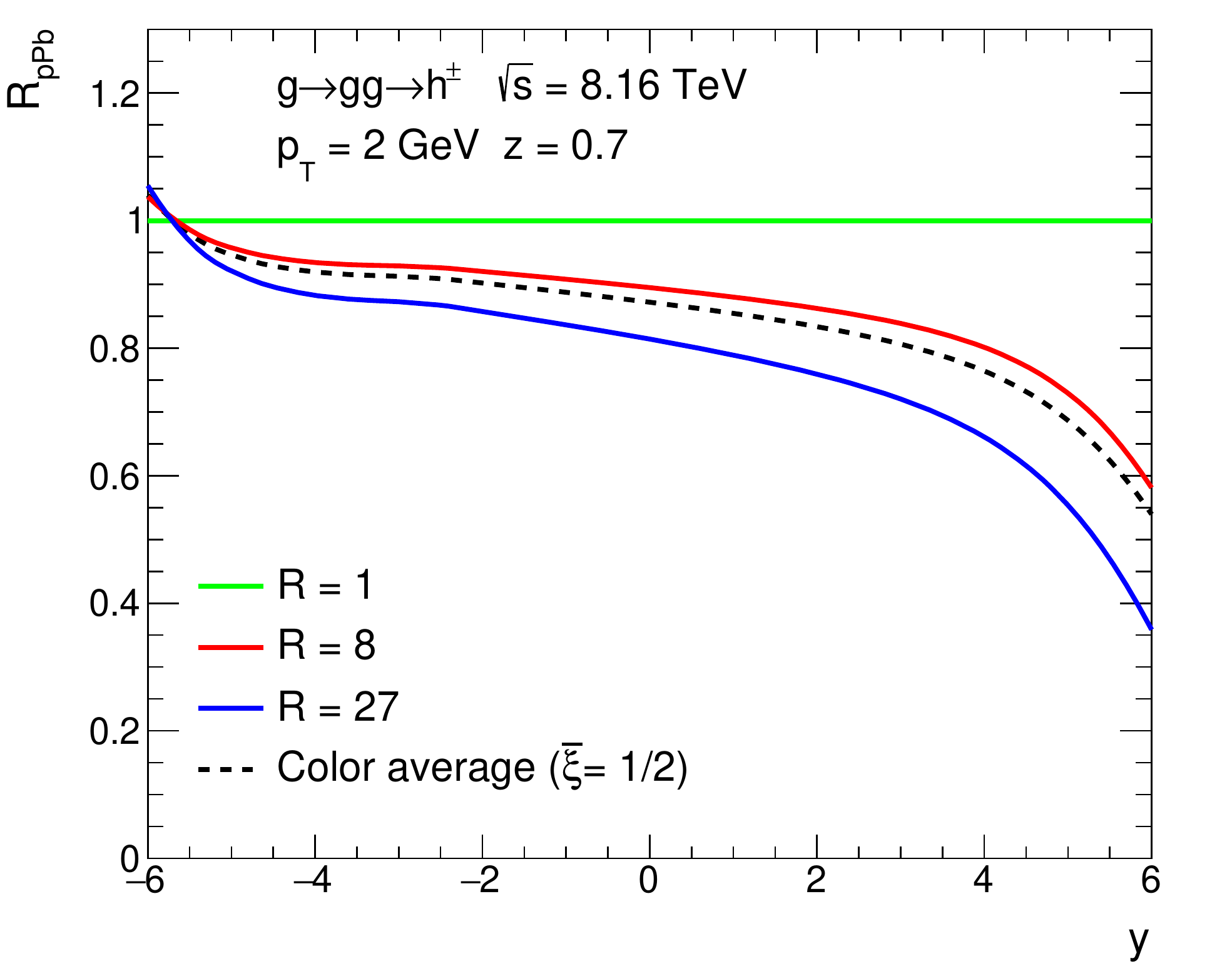}
\includegraphics[width=7.5cm]{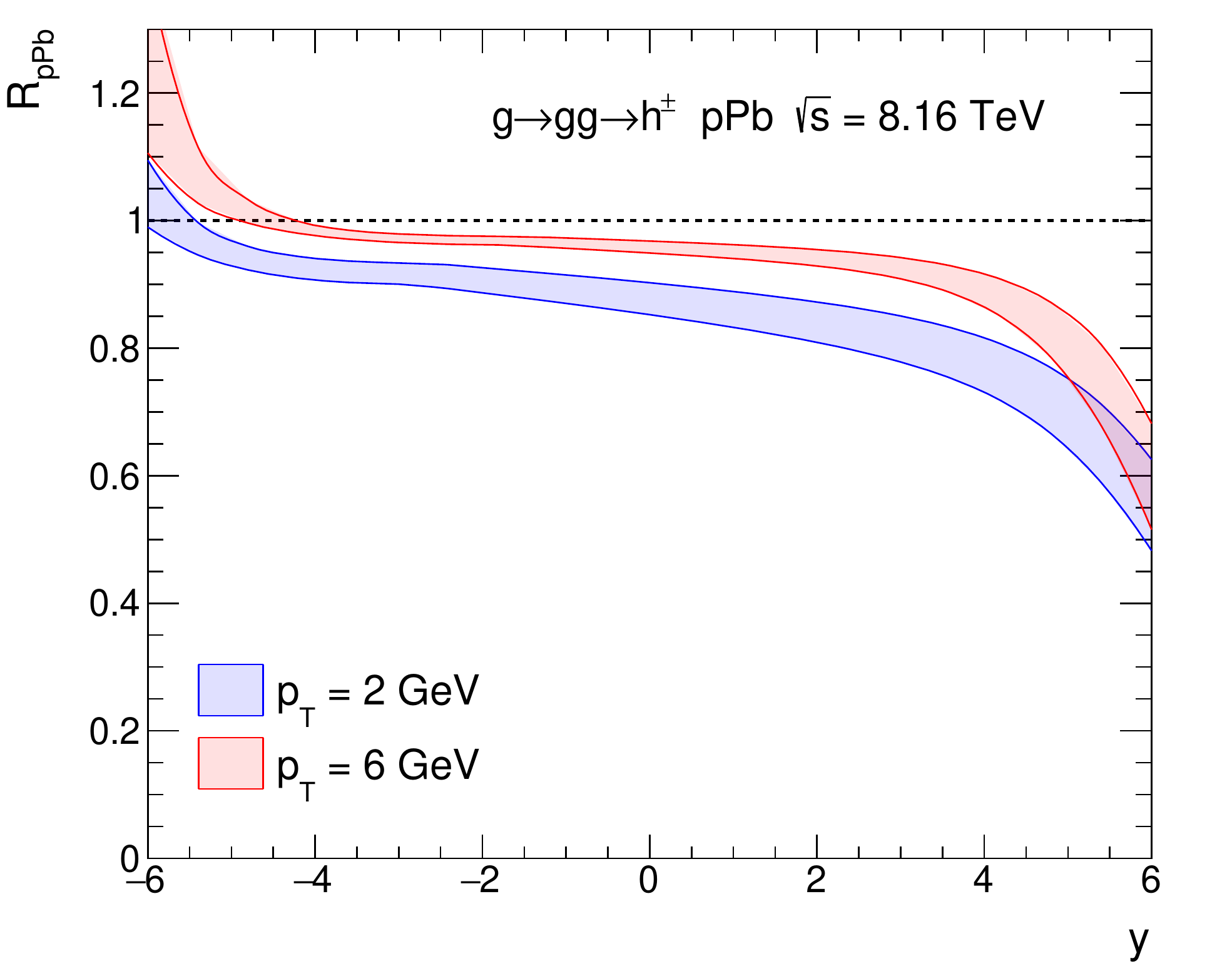}  
\caption{Left: Rapidity dependence of  $R_{\pA}^{\R}$, Eq.~\eq{RpA-yR}, at $\pt = 2$~GeV, in the $g\to (gg)_{_\R}$ channel for $\R={\bf 1}$ (green line), $\R={\bf 8}$ (red line), and $\R={\bf 27}$ (blue line). The color-averaged nuclear modification factor $R_{\pA}^{h}$, Eq.~\eq{RpA-y-master}, is shown for $\bar{\xi}=1/2$ (dashed black line). Right: $R_{\pA}^{h}$ as a function of $y$ and associated uncertainty band, for different values of $\pt$. Calculations are done at $\sqrt{s}=8.16$~TeV in the $g\to gg$ channel.}
\label{fig-RpA-gg}
\end{figure}

In Fig.~\ref{fig-RpA-gg} (left) we show the rapidity dependence of $R_{\pA}^{\R}$ at fixed $\pt=2$~GeV, for the three color states $\R={\bf 1}, {\bf 8}, {\bf 27}$.  (At leading-order, the probability for the $gg$ pair to be in the decuplet state $\R={\bf 10 \oplus \overline{10}}$ vanishes, see Appendix~\ref{app-g2gg-case}.) When the $gg$ final state is color singlet, the fully coherent induced gluon spectrum vanishes, hence no FCEL effect is expected in this case, $R_{\pA}^{{\bf 1}}=1$. More interesting is the suppression of the octet $gg$ final state, $R_{\pA}^{\bf 8}<1$, which is predicted in the entire rapidity range considered here, $-6<y<6$. The shape is reminiscent of the suppression predicted for quarkonium in Ref.~\cite{Arleo:2012rs}. At $y=0$, the nuclear production ratio is $R_{\pA}^{{\bf 8}}\simeq 0.9$, while the suppression is stronger at larger rapidity, $R_{\pA}^{\bf 8}\simeq 0.6$ at $y=6$, due to the \pp cross section steeply falling at large $y$. Note that the steeply {\it rising} \pp cross section at very backward rapidity leads to a slight enhancement $R_{\pA}^{\bf 8}>1$ below $y\simeq-6$. For the 27-plet $gg$ final state, the suppression expected from FCEL follows the same pattern, but is more pronounced than for the octet state, due to the larger Casimir, $C_{27}=2\,(\Nc+1)$, in the prefactor \eqref{color-rule} of the induced gluon spectrum. Shown as a dashed line is the color-averaged nuclear production ratio \eqref{RpA-y-master} (obtained using the probabilities given in \eq{color-proba-pheno-gg}),\footnote{It turns out to be numerically very close to $R_{\pA}^{\bf 8}$, see Fig.~\ref{fig-RpA-gg} (left). An even more striking coincidence appears in the case of the $q \to qg$ underlying process, see section \ref{app:other-processes}, between the color-averaged modification factor and the color state $\R={\bf \bar{6}}$ of the final $qg$ pair, see Fig.~\ref{fig-RpA-qg} (left).} which we now discuss in more detail.

\begin{figure}[t]
\centering
\includegraphics[width=7.5cm]{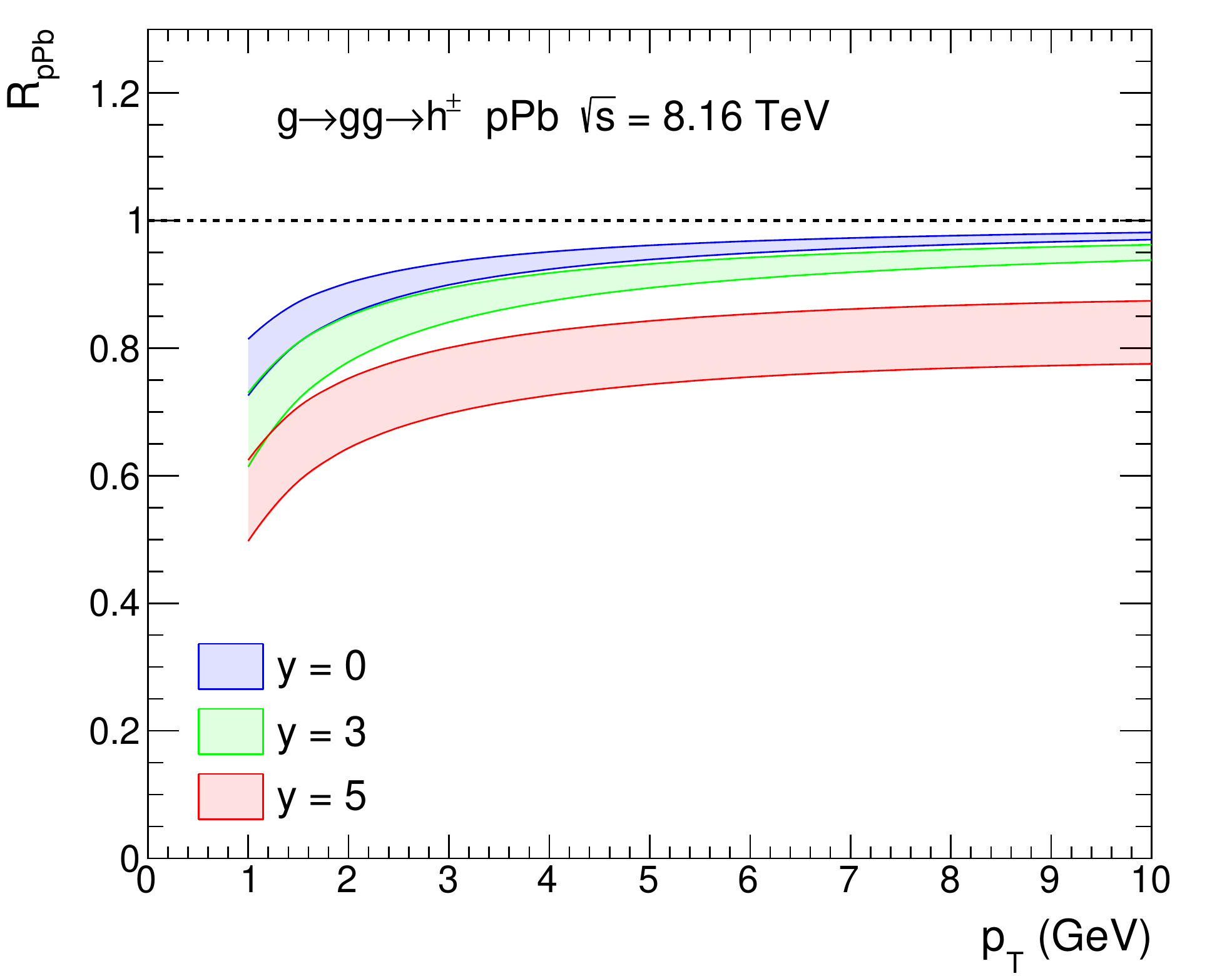}  
\caption{Nuclear modification factor \eq{RpA-y-master} as a function of $\pt$ for different values of $y$. Calculations are done at $\sqrt{s}=8.16$~TeV in the $g\to gg$ channel.}
\label{fig-RPA-pt}
\end{figure}

The light hadron suppression is shown in Fig.~\ref{fig-RpA-gg} (right) as a function of $y$. The shape of $R_{\pA}^{h}$ at $\pt=2$~GeV is discussed above. Because of the specific dependence of the induced spectrum \eq{dIR} in $K_{_\perp} = \pt / z$ (recall that in the PDA we have $M_{_\perp} \to M_{\bar{\xi}} = K_{_\perp}/\sqrt{\bar{\xi}(1-\bar{\xi})}$ in \eq{dIR}--\eq{quenching-dilog}, see section~\ref{sec:rescaling}), FCEL effects weaken at larger $\pt$, as can be seen when comparing the predictions at $\pt=2$~GeV and $\pt=6$~GeV. In the latter case, the shape is similar and the suppression is more moderate, except at very large rapidity, $y \gtrsim 5$, where the effects of the slope of the cross section are larger at higher $\pt$ due to the more restricted phase space. Similarly, the enhancement already mentioned at very backward rapidity is now clearly visible below $y\simeq-5$. 
The $\pt$ dependence of the nuclear production ratio shown in Fig.~\ref{fig-RPA-pt} can be simply understood, as $R_{\pA}^{h}$ approaches unity at large $\pt$ due to the scale dependence of the induced gluon spectrum. At $y=5$ however, the suppression appears to flatten for $\pt > 5$~GeV, the effect of the scale dependence being compensated by the strong phase space restriction when the rapidity and transverse momentum are both large. 

\begin{figure}[t]
\centering
\includegraphics[width=7.5cm]{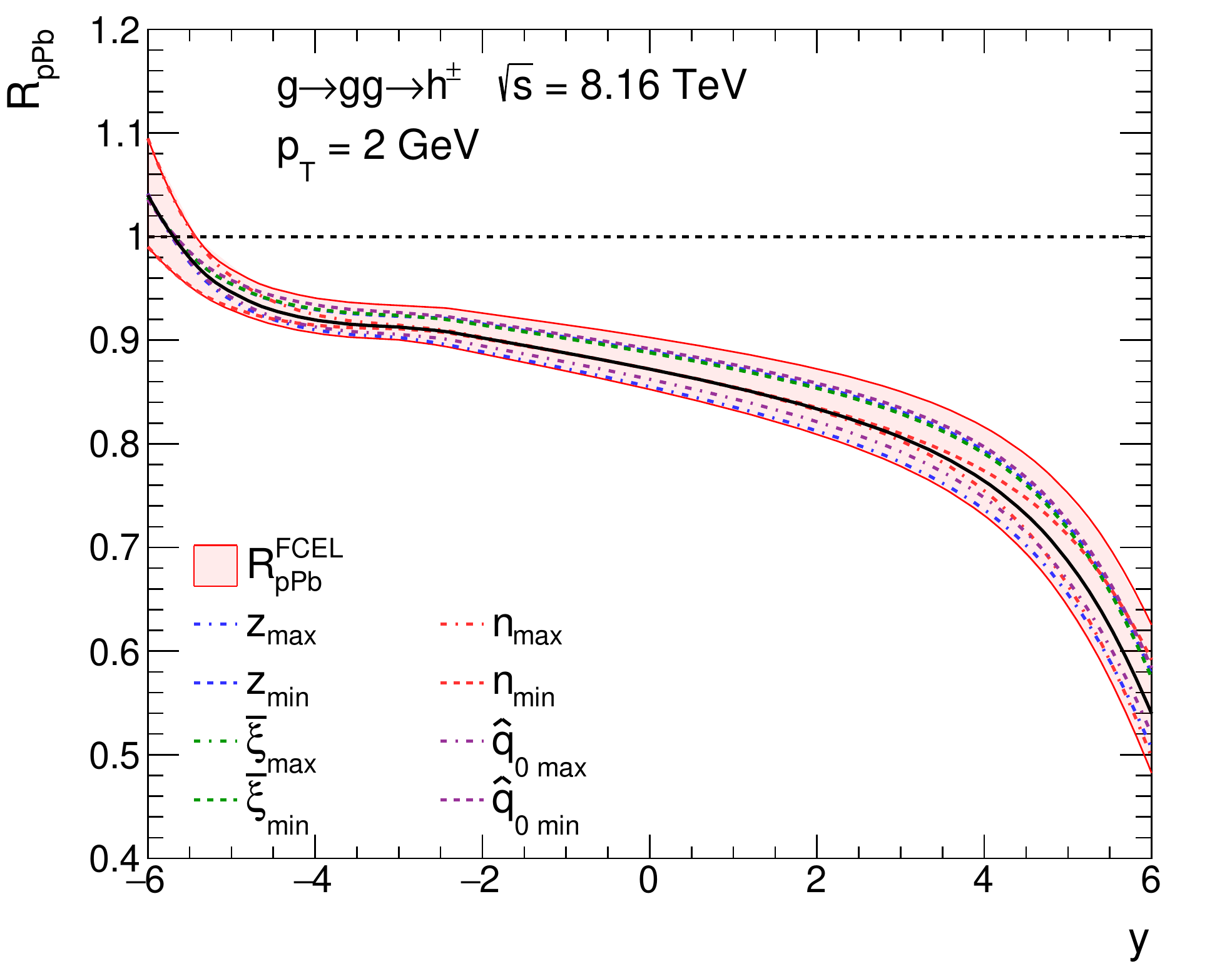}
\includegraphics[width=7.5cm]{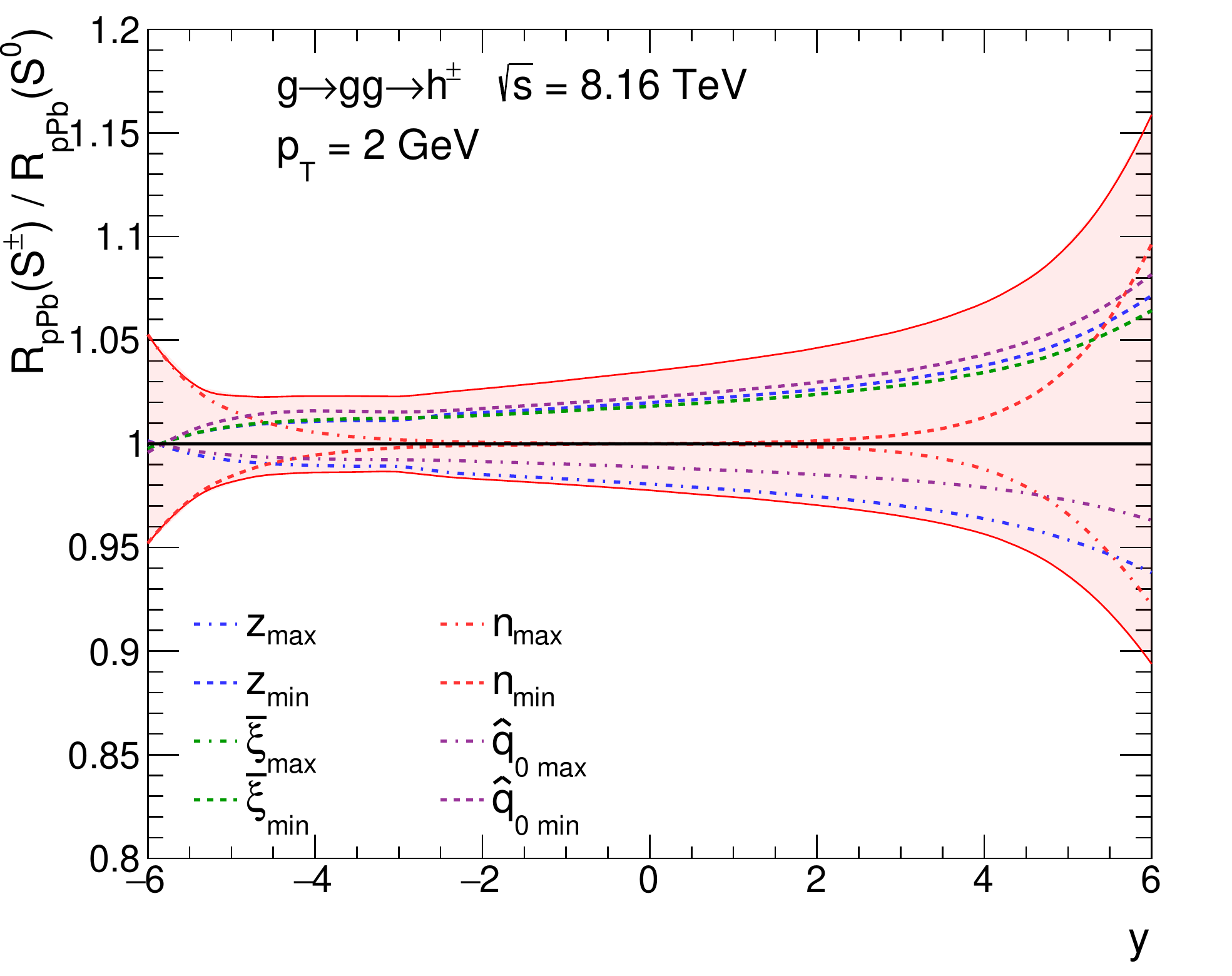}
\caption{Left: Individual contributions $R_{\pA}(S_k^{\pm})$ to the uncertainty of  $R_{\pA}$ at $\pt = 2$~GeV, in the $g\to gg$ channel. Calculations are shown for maximal (dash-dotted lines) and minimal (dashed lines) values of $z$ (blue), $\bar{\xi}$ (green), $n$ (red), and $\hat{q}_{_0}$ (purple). The filled band corresponds to the total uncertainty defined by \eqref{eq:errors}. Right: Relative uncertainty $R_{\pA}(S_k^{\pm})/R_{\pA}(S_{}^{0})$.}
\label{fig-uncertainties}
\end{figure}

We now discuss the theoretical uncertainties. As discussed in section~\ref{sec:procedure}, the uncertainty of our predictions comes from the independent variations of $n$, $\bar{\xi}$, $z$ and $\hat{q}_{_0}$, resulting in the bands shown in Fig.~\ref{fig-RpA-gg} (right) and Fig.~\ref{fig-RPA-pt}. 
In order to give a feeling on the contributions from each parameter variation to the total uncertainty, the curves for $R_{\pA}(S^+_k)$ (dash-dotted) and $R_{\pA}(S^-_k)$ (dashed) are shown individually in Fig.~\ref{fig-uncertainties} (left), as a function of rapidity at fixed $\pt=2$~GeV. In most of the rapidity range, the upper uncertainty on $R_{\pA}^h$ from the (lower) variation of $z$, $\bar{\xi}$ and $\hat{q}_{_0}$ has a similar magnitude, while the variation of $n$ only contributes to the total uncertainty at very forward/backward rapidities, $|y|\gtrsim5$. The lower uncertainty of $R_{\pA}^h$ is dominated by the (upper) variation of $z$, except again at large $|y|$ where the influence of $n$ becomes the largest. This can also be seen in Fig.~\ref{fig-uncertainties} (right) showing the individual ratios $R_{\pA}(S^\pm_k)/R_{\pA}(S^0)$ quantifying the relative uncertainty. Remarkably, the uncertainty remains very small, at the level of a few percent around mid-rapidity (and at most 15\% at the largest rapidity considered, $y=6$). 

We stress that the smallness of FCEL relative uncertainties is expected within our approach, since FCEL is fully determined within perturbative QCD. Moreover, the parameters $\bar{\xi}$, $z$, and $\hat{q}_{_0}$ enter the induced gluon spectrum \eq{dIR} only through the product $\hat{q}_{_0}\,z^2\,\bar{\xi} (1-\bar{\xi})$ in the argument of a logarithm. This logarithm turns out to be large, consistently with the accuracy \eq{log-accuracy} of our approach. Varying the parameters by about $\pm 50\%$ therefore leads to variations which are formally beyond the leading logarithm, resulting in a narrow uncertainty band. 
As for the parameter $n$, its variation affects negligibly the predictions, except at very large $|y|$ where it dominates the total relative uncertainty (see Fig.~\ref{fig-uncertainties}), which however remains moderate. 

\subsection{Comparison to data}
\label{sec:predictions-vs-data}

Let us now compare the FCEL expectations to the measurements of light hadron suppression in pPb collisions at $\sqrt{s}=5.02$~TeV by the ALICE experiment~\cite{Adam:2016dau,Acharya:2018hzf}.

\begin{figure}[t]
\centering
\includegraphics[width=7.5cm]{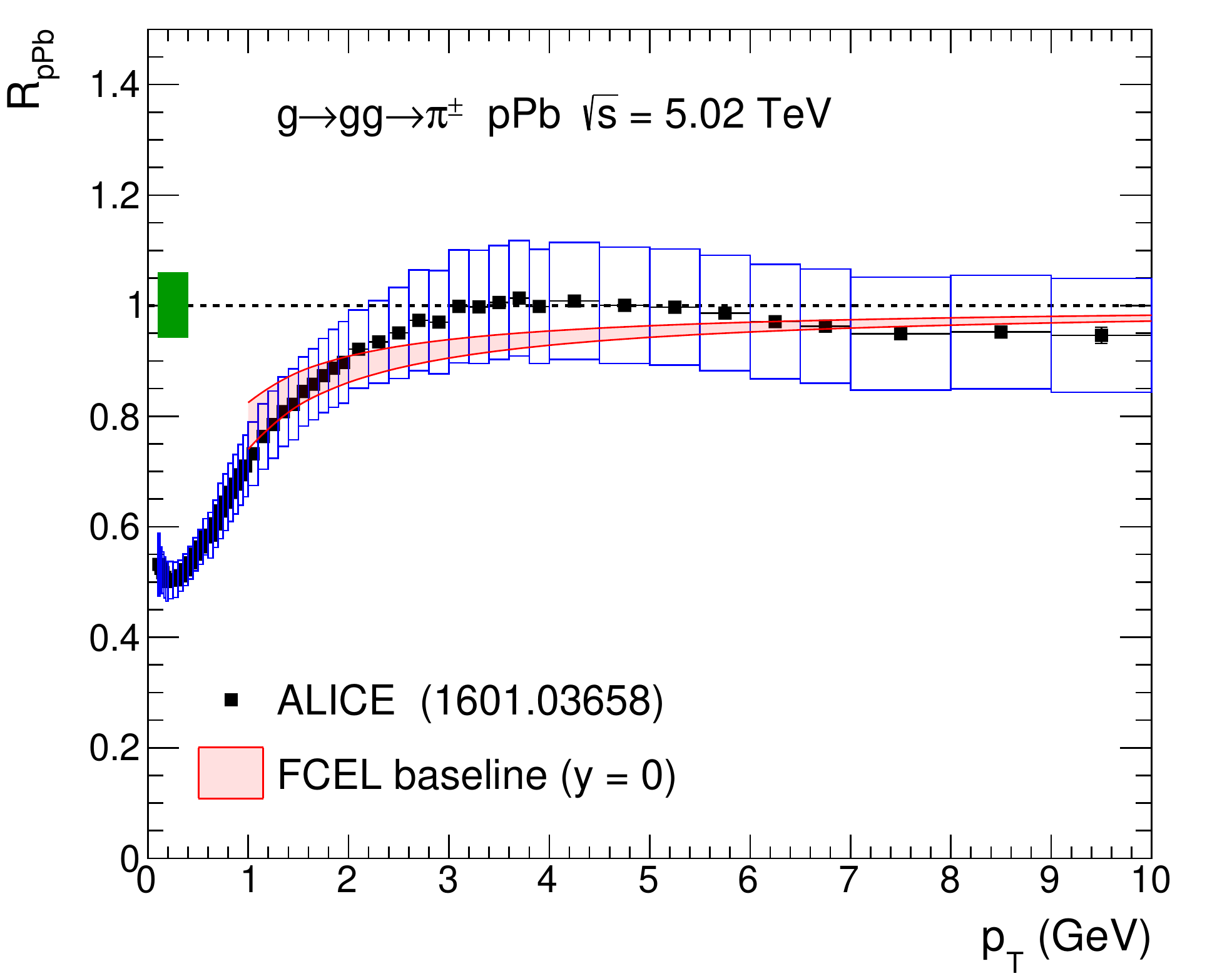} 
\includegraphics[width=7.5cm]{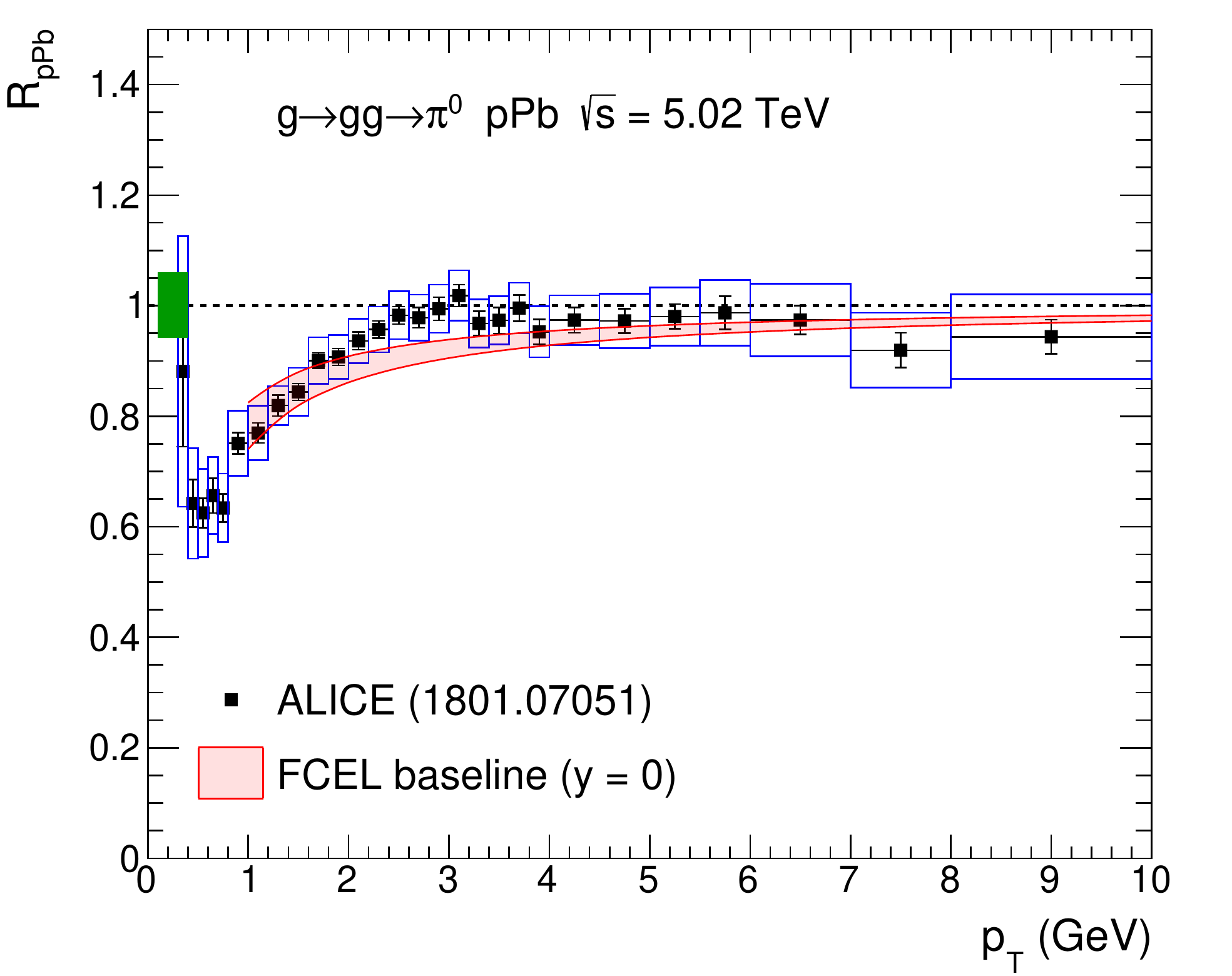} \\
\includegraphics[width=7.5cm]{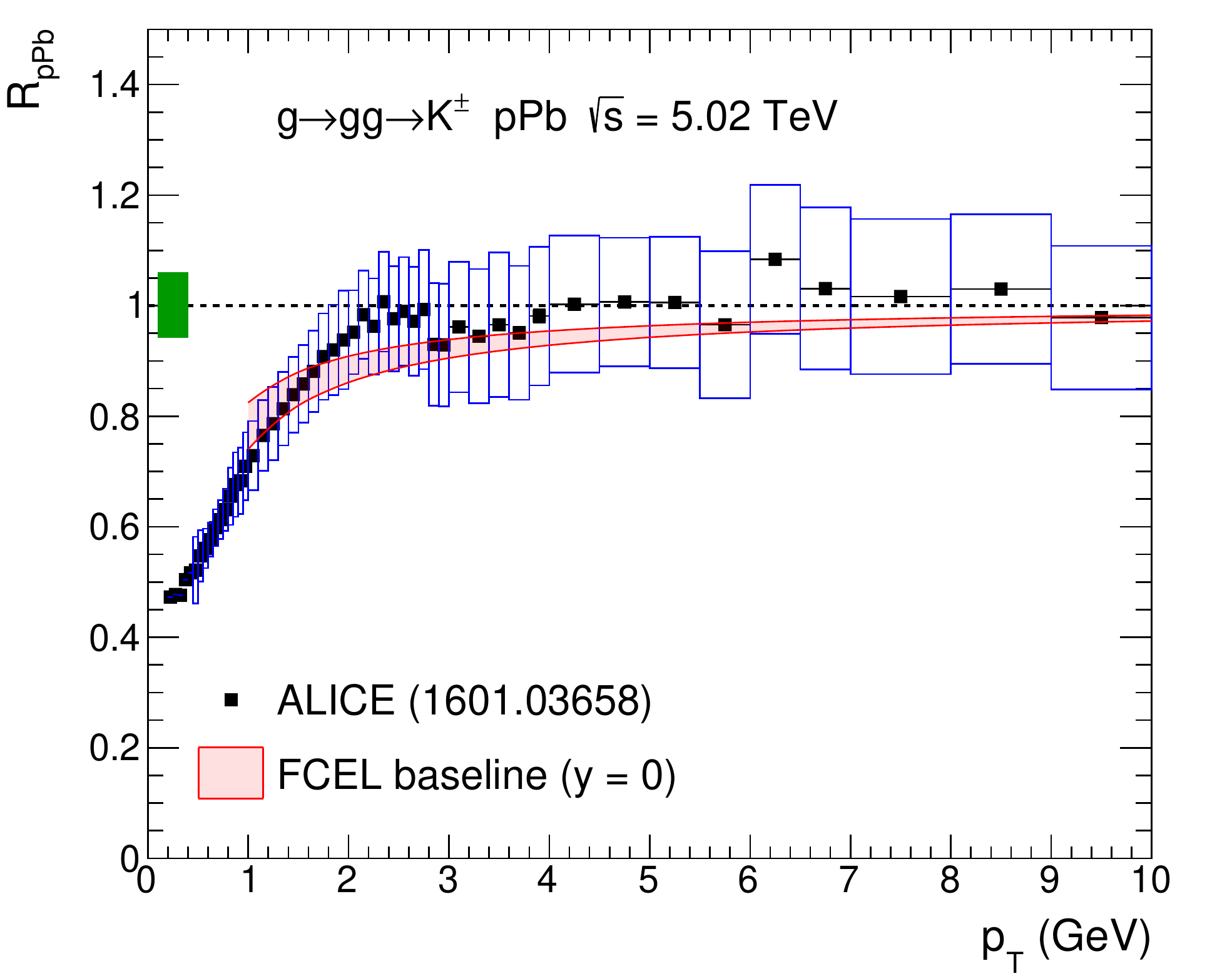}  
\includegraphics[width=7.5cm]{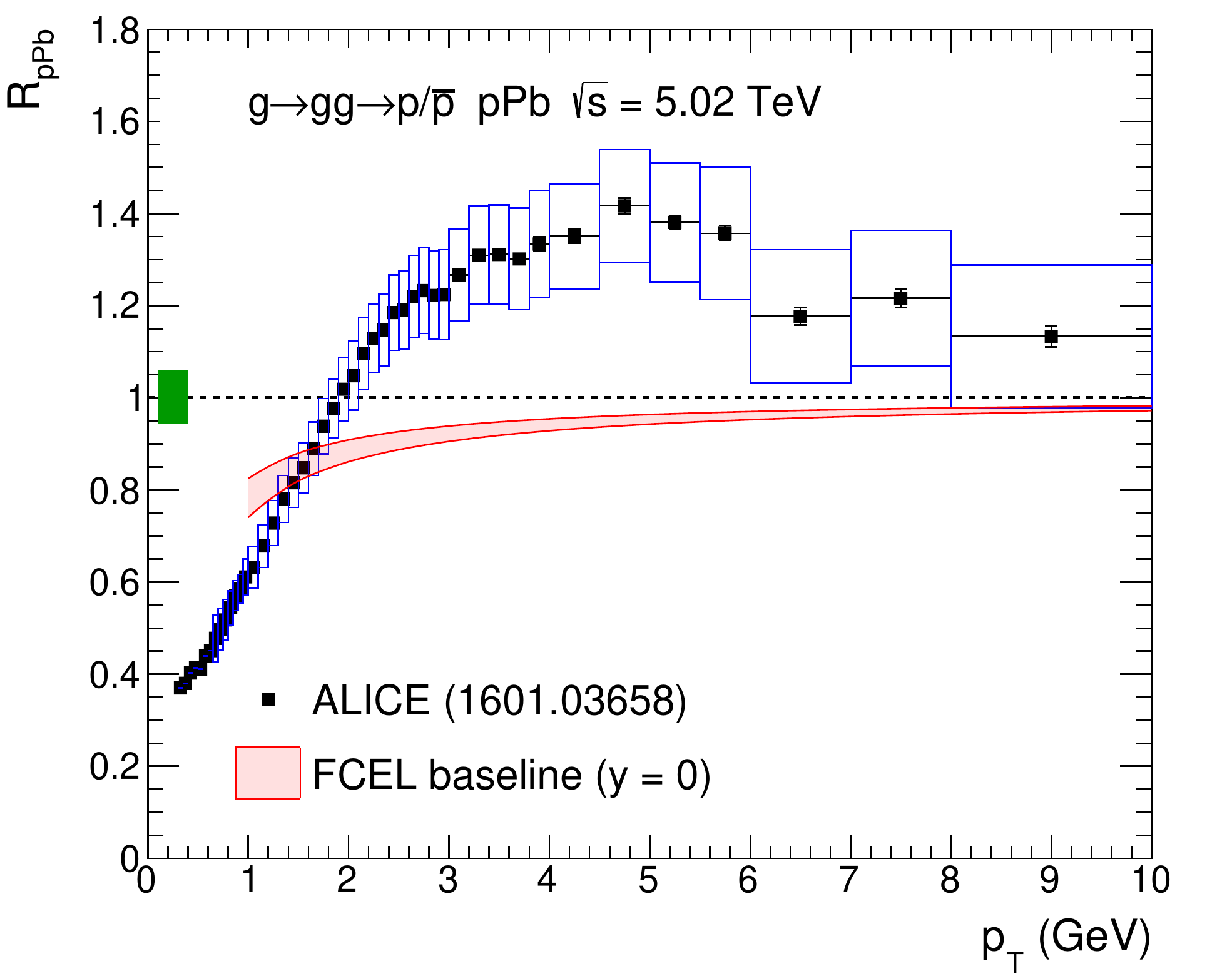} 
\caption{Nuclear modification factor \eq{RpA-y-master} in pPb collisions at $\sqrt{s}=5.02$~TeV (red band) in comparison to ALICE charged pion data (top left)~\cite{Adam:2016dau}, neutral pion data (top right)~\cite{Acharya:2018hzf}, charged kaon data (bottom left)~\cite{Adam:2016dau}, and proton/antiproton data (bottom right)~\cite{Adam:2016dau}. The baseline calculation assuming FCEL effects only is done in the $g\to gg$ channel and at $y=0$.}
\label{fig-RPA-data}
\end{figure}

The measured nuclear production ratio is shown in Fig.~\ref{fig-RPA-data} for $\pi^\pm$ (top left), $\pi^0$ (top right), $K^\pm$ (bottom left), and $p/\bar{p}$ (bottom right) production, as a function of $\pt$. In each panel of Fig.~\ref{fig-RPA-data}, we superimpose our prediction for the nuclear production ratio expected from the sole FCEL effect.\footnote{This prediction is shown for $\pt>1$~GeV to ensure a perturbative picture ($K_{_\perp} = \pt / z \gg \Lambda_{_{\rm QCD}}$).} Let us note that in our approach, the suppression of different (light) hadron species is expected to be similar. Indeed, although some differences could arise from a significantly different shape of the pp cross section or a different fragmentation pattern, such differences should lie within the theoretical uncertainty band which includes the uncertainties associated with the variation of $n$ and $z$. 

This {\it baseline} prediction --~assuming only FCEL effects~-- proves to be in good agreement with ALICE measurements of pions and kaons.\footnote{A similar agreement appears between the `FCEL baseline' and CMS charged hadron data~\cite{Khachatryan:2015xaa}.} In particular, the data sometimes attributed to saturation~\cite{Tribedy:2011aa,Albacete:2012xq,Rezaeian:2012ye,Lappi:2013zma}, nPDF effects~\cite{Helenius:2012wd}, and Cronin effect with initial-state energy loss~\cite{Kang:2012kc}, are well reproduced here by fully coherent energy loss alone. Since the uncertainties of the FCEL baseline  prediction are small (as discussed in the end of section~\ref{sec:LHC-pheno}) and significantly smaller than those of the measurements, taking into account FCEL should therefore provide strict constraints on other physical processes. Unlike meson production, FCEL alone is clearly not sufficient to explain proton/antiproton data, which exhibit a significant enhancement reminiscent of the Cronin effect above $\pt\gtrsim2$~GeV. 

It would be interesting to compare the present baseline predictions at large rapidity, where FCEL effects become the strongest (see Fig.~\ref{fig-RpA-gg}, right). Such measurements could be performed in the near future by the LHCb experiment in the rapidity region $2 < y < 4.5$~\cite{Aaij:2014pza}. Measurements at even larger rapidities by the LHCf experiment, $y > 8.8$, should also be sensitive to FCEL, although the $\pt$ coverage in the present data, $\pt\lesssim0.8$~GeV~\cite{Adriani:2015iwv}, is limited and barely perturbative.

\section{Other partonic subprocesses}
\label{app:other-processes}

The results discussed in section~\ref{sec:predictions} have been obtained assuming that hadron production in \pp collisions at LHC is dominated by $g\to gg$ forward scattering (see Fig.~\ref{fig:typical-process}), which as mentioned in section~\ref{sec:subprocesses} is a sensible assumption at mid-rapidity. Here we present the FCEL baseline predictions obtained assuming $q \to qg$ (section~\ref{sec:qg}) and $g \to q \bar{q}$ (section~\ref{sec:qqb}). As we shall see, these predictions are qualitatively similar to those obtained for $g \to gg$. 

\subsection{{\texorpdfstring{${q\to qg}$}{}}}
\label{sec:qg}

As for $g\to gg$, the light hadron suppression in the $q\to qg$ channel is computed from \eqref{RpA-y-master}, but with now  the accessible color states of the final $qg$ pair being $\R= {\bf 3}, {\bf \bar{6}}, {\bf 15}$. The corresponding probabilities $\rho_{_\R}(\xi)$ are given in~\eq{color-proba-pheno-qg} of Appendix \ref{app:color-proba}. 

\begin{figure}[t]
\centering
\includegraphics[width=7.5cm]{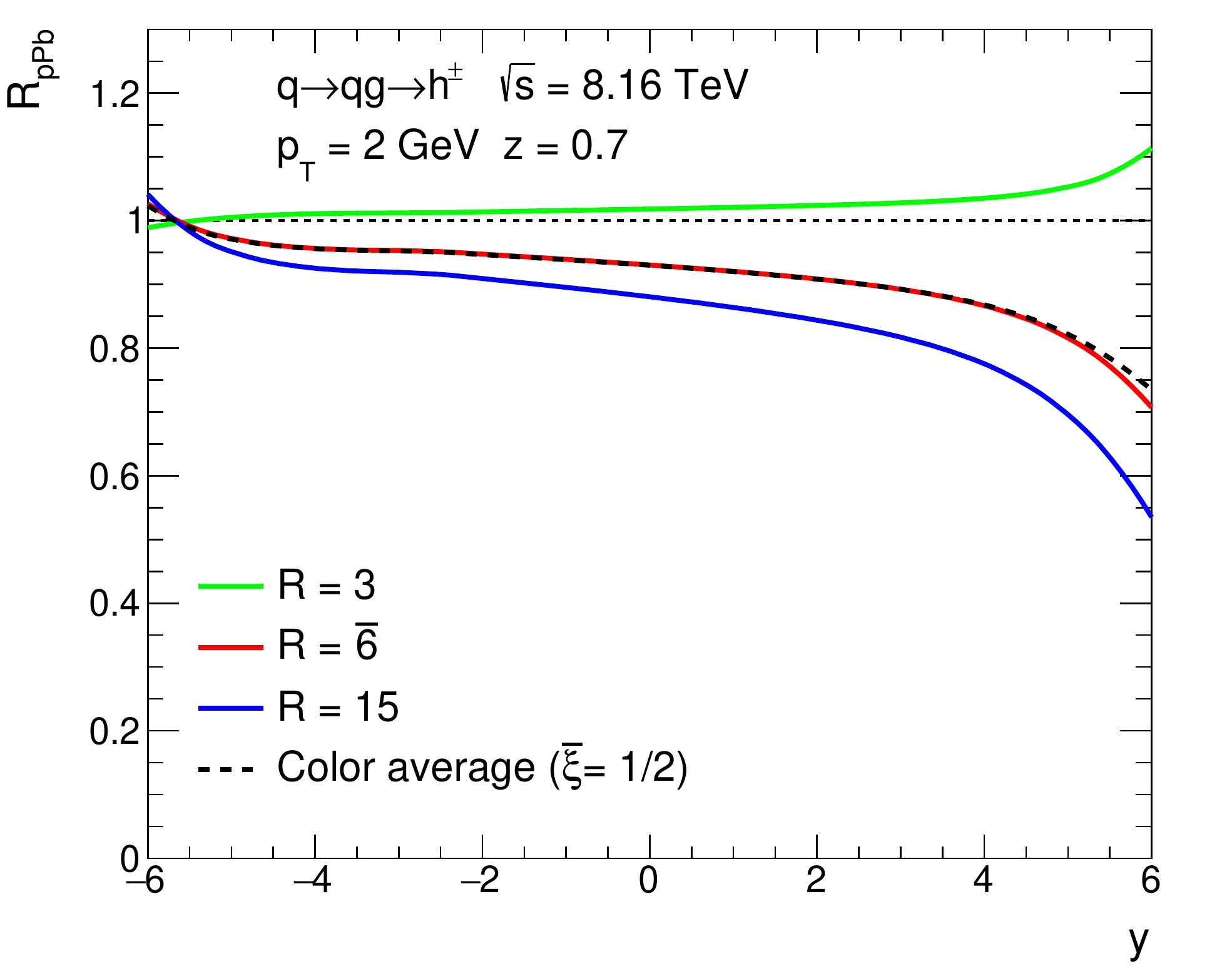}
\includegraphics[width=7.5cm]{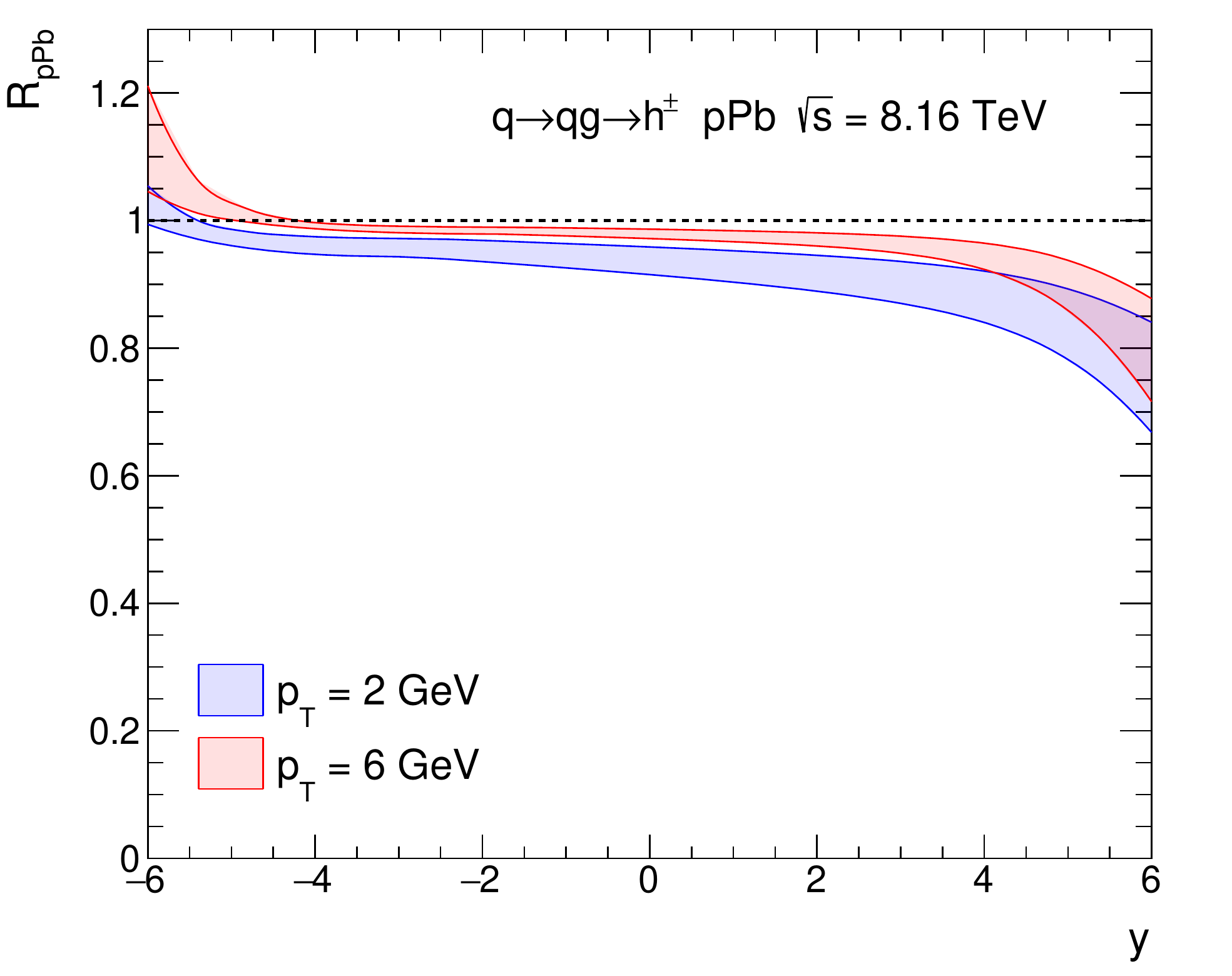} 
\caption{Left: Rapidity dependence of  $R_{\pA}^{\R}$ at $\pt = 2$~GeV, in the $q\to (qg)_{_\R}$ channel for $\R={\bf 3}$ (green line), $\R={\bf \bar{6}}$ (red line), $\R={\bf 15}$ (blue line). The color-averaged nuclear production ratio, $R_{\pA}^{h}$, is shown for 
$\bar{\xi}=1/2$ (dashed black line). Right: Rapidity dependence of $R_{\pA}^{h}$ at $\pt=2$~GeV (blue band) and $\pt=6$~GeV (red band) in the $q\to qg$ channel.}
\label{fig-RpA-qg}
\end{figure}

The rapidity dependence of $R_{\pA}^{\R}$ is shown in Fig.~\ref{fig-RpA-qg} (left) at fixed $\pt=2$~GeV, for the three color states. The shape of $R_{\pA}^{\,{\bf \bar{6}}}$ and $R_{\pA}^{{\bf 15}}$ is similar to that of $R_{\pA}^{\,{\bf 8}}$ and $R_{\pA}^{{\bf 27}}$ in the $g\to gg$ channel, see Fig.~\ref{fig-RpA-gg} (left). The suppression is however less pronounced due to smaller color prefactors~\eqref{color-rule} in the induced gluon spectrum, $F_c^{{\bf \bar{6}}} /F_c^{{\bf 8}}= 5/9$ and $F_c^{{\bf 15}} /F_c^{{\bf 27}} = 11/24$. 
As mentioned after Eq.~\eq{RpA-yR-gain}, when the final $qg$ pair is a color triplet, $\R={\bf 3}$, lesser (fully coherent) radiation is expected in pA with respect to pp collisions, due to a negative color prefactor in this case. This induced {\it energy gain} leads to $R_{\pA}^{{\bf 3}}>1$, as can be seen in Fig.~\ref{fig-RpA-qg} (left, green solid line). This enhancement is, however, modest since the color prefactor is small, $F_c=-1/\Nc$. It becomes more pronounced at larger rapidity, due to the steepness of the pp cross section: even a small energy gain shifting the rapidity from $y$ to $y - \delta < y$ in Eq.~\eq{RpA-yR-gain} can lead to a significant enhancement.

The uncertainty band for the `inclusive' nuclear production ratio $R_{\pA}^h$, obtained using the procedure described in section~\ref{sec:procedure}, is shown in Fig.~\ref{fig-RpA-qg} (right) for $\pt=2$~GeV and $\pt=6$~GeV. It is qualitatively similar to that in the $g\to gg$ channel, see Fig.~\ref{fig-RpA-gg} (right). 

\subsection{{\texorpdfstring{${g\to q\bar{q}}$}{}}}
\label{sec:qqb}

In the $g\to q\bar{q}$ channel, the possible $q\bar{q}$ color states are $\R= {\bf 1}, {\bf 8}$. The probabilities $\rho_{_\R}(\xi)$ to be used in \eqref{RpA-y-master} are given in Eq.~\eq{color-proba-qqbar}. 

\begin{figure}[t]
\centering
\includegraphics[width=7.5cm]{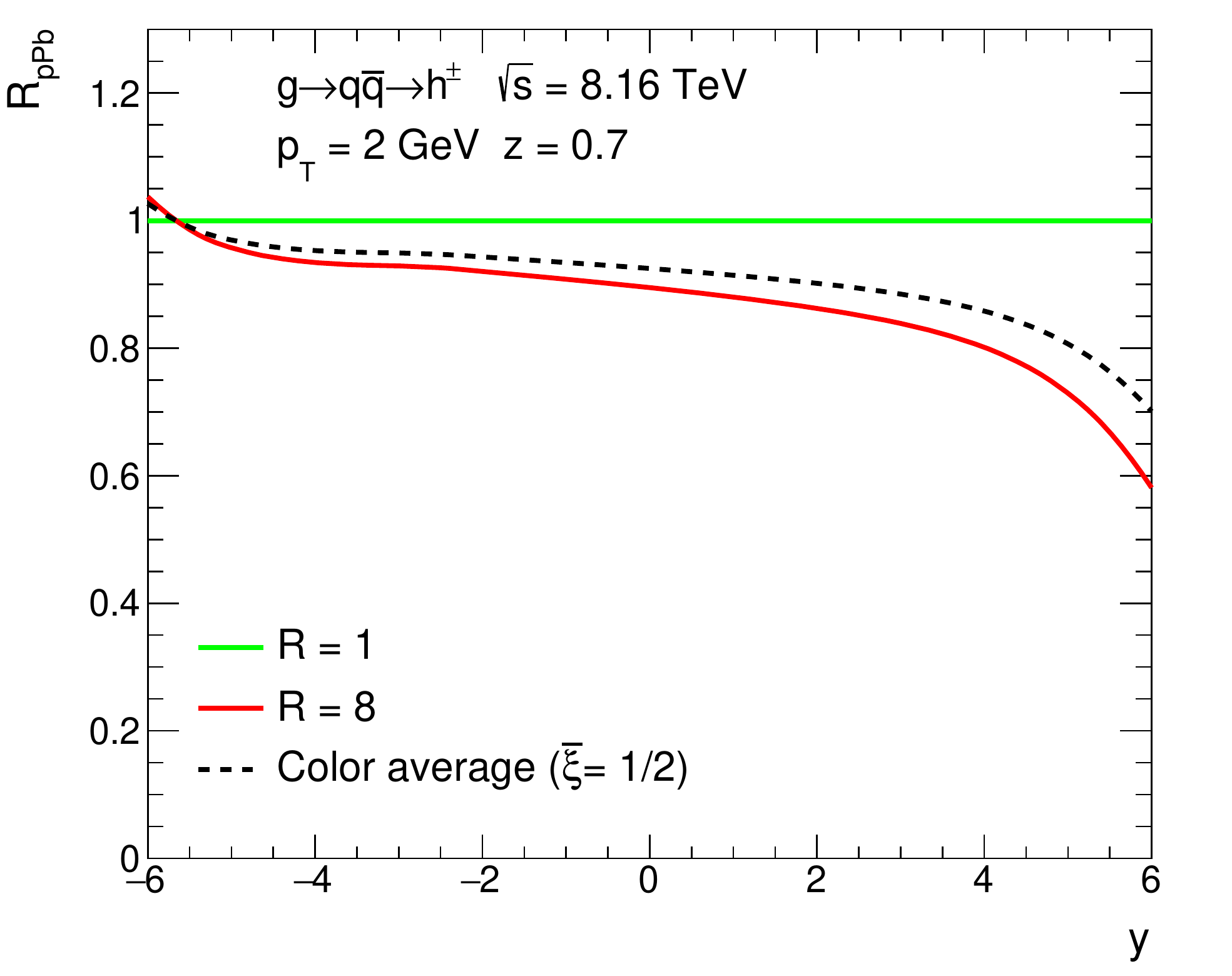}
\includegraphics[width=7.5cm]{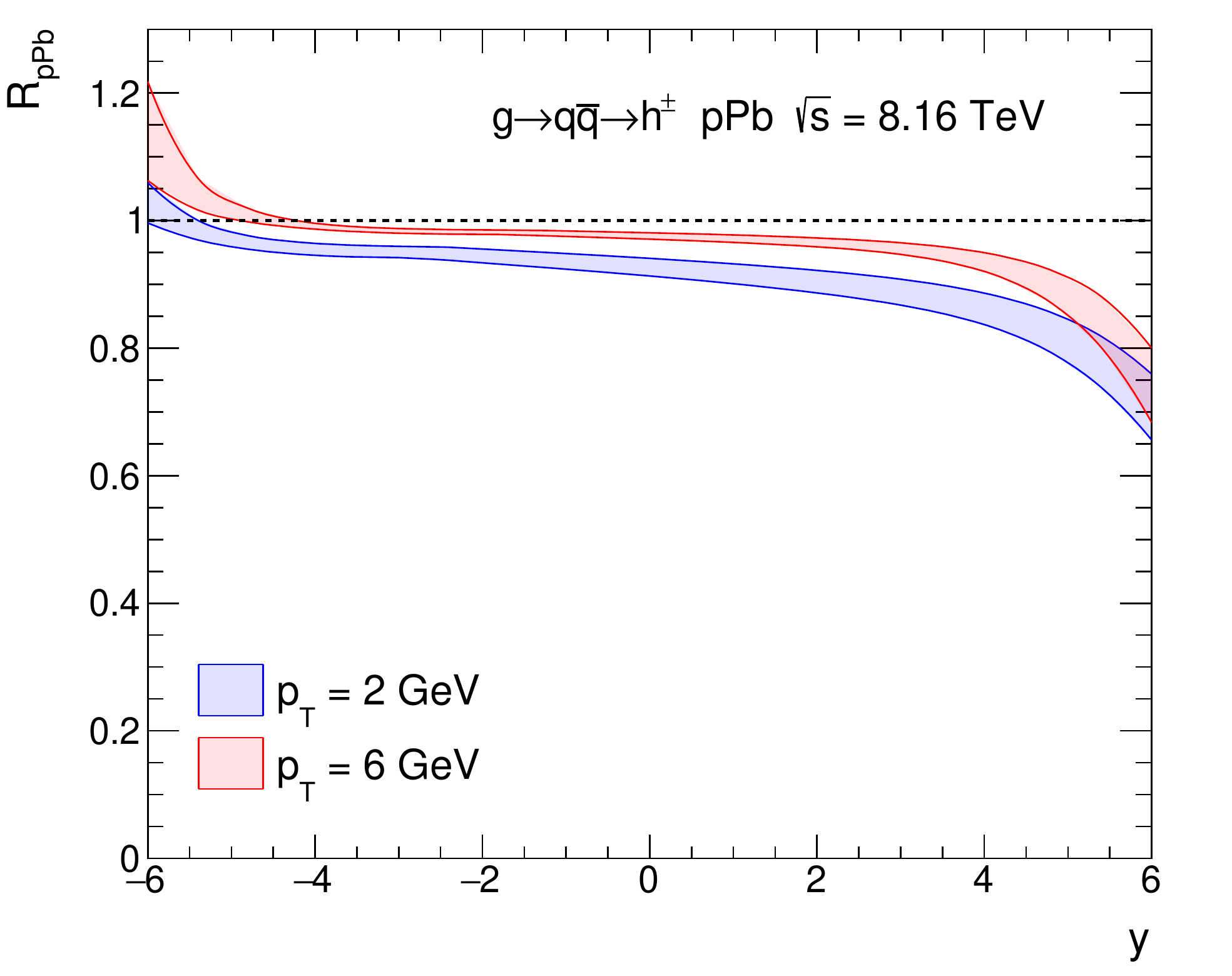} 
\caption{Same as Fig.~\ref{fig-RpA-qg} in the $g \to q\bar{q}$ channel ($\R={\bf 1}, {\bf 8}$).}
\label{fig-RpA-qqb}
\end{figure}

The nuclear production ratios $R_{\pA}^{\R}$ for $\R= {\bf 1}, {\bf 8}$ are shown in Fig.~\ref{fig-RpA-qqb} (left). They obviously coincide with those obtained for the same final color states in the $g\to gg$ channel, see Fig.~\ref{fig-RpA-gg} (left). The difference between the $g\to q\bar{q}$ and $g\to gg$ channels appears for the color-averaged nuclear production ratio $R_{\pA}^h$, see the dashed lines in Fig.~\ref{fig-RpA-gg} (left) and Fig.~\ref{fig-RpA-qqb} (left), since $\R={\bf 27}$ is only accessible in the latter channel (and the probabilities $\rho_{_\R}(\xi)$ for $\R= {\bf 1}, {\bf 8}$ are also different in the two channels). As a consequence, the resulting hadron suppression is slightly less pronounced in the $g\to q\bar{q}$ channel. 

In Fig.~\ref{fig-RpA-qqb} (right) we show $R_{\pA}^h$ as a function of $y$, for $\pt=2$~GeV and $\pt=6$~GeV, after taking into account theoretical uncertainties. We see that $R_{\pA}^h$ exhibits the same characteristic shape as in the $g\to gg$ and $q \to qg$ channels. Surprisingly, the uncertainty band in Fig.~\ref{fig-RpA-qqb} (right) proves narrower than in the other channels, which is due to a numerical coincidence. When $\bar{\xi}$ deviates from its central value $\bar{\xi} = 0.5$, both the $q\bar{q}$ invariant mass~\eqref{dijet-mass} and effective color charge (see Fig.~\ref{fig-probas}, right) increase, the former effect leading to lesser suppression, the latter to stronger suppression. It turns out that these two effects almost perfectly balance in the $g\to q\bar{q}$ channel, making $R_{\pA}^h$ almost independent of $\bar{\xi}$, resulting in a reduced overall theoretical uncertainty.

\section{Discussion and outlook}
\label{sec:discussion}

The production of all hadron species in pA collisions is expected to be affected by FCEL~\cite{Arleo:2010rb}. In this work, calculations of the light hadron nuclear production ratio are provided in pPb collisions at top LHC energy, $\sqrt{s}=8.16$~TeV, in order to provide FCEL baseline predictions taking only this effect into account. Interestingly, the predictions at $y=0$ (for $g \to gg$ forward scattering) agree well with the $\pt$ dependence of $R_{\pA}$ of mesons measured by ALICE at $\sqrt{s}=5.02$~TeV (while a significant deviation is observed in the case of $p/\bar{p}$ production).  

The suppression due to FCEL alone appears to be of the same order of magnitude as nPDF~\cite{Helenius:2012wd} or saturation~\cite{Tribedy:2011aa,Albacete:2012xq,Rezaeian:2012ye,Lappi:2013zma} effects at mid-rapidity. These observations indicate that FCEL should be taken into account in phenomenological interpretations of the pA data, and in particular in nPDF global fit analyses using hadron production data in pA collisions. All the more so as the uncertainties of the FCEL baseline predictions are very small, only 4\% at mid-rapidity (see section~\ref{sec:LHC-pheno}), much smaller than the uncertainty of the predictions of nPDF or saturation effects. As an illustration, we have estimated the nPDF relative uncertainty at mid-rapidity, using nCTEQ15~\cite{Kovarik:2015cma} and EPPS16~\cite{Eskola:2016oht} nPDF sets, to be 25\% and 40\% at $\pt=2$~GeV, respectively. Note that the smallness of FCEL uncertainties is expected from the variation of parameters affecting the induced gluon spectrum beyond the leading logarithm. 

In order to properly predict FCEL effects away from the mid-rapidity region, one would need to consider other partonic processes eventually overcoming the $g \to gg$ process considered in section~\ref{sec:predictions}. A first step in this direction has been made in section~\ref{app:other-processes}, where we computed the nuclear production ratio from FCEL effects in the $g \to q \bar{q}$ and $q \to qg$ channels, the latter being particularly relevant for hadron production at large forward rapidity. Note that the $a \to bc$ forward scatterings considered in our study correspond to the $2 \to 2$ partonic processes $a + g_{_\textnormal{A}} \to bc$ when viewed in the proton-nucleon c.m.~frame, $g_{_\textnormal{A}}$ denoting the gluon from the target nucleus. The three channels considered should thus be relevant when the nucleus content is dominated by gluons, \ie, at small $\xtwo$, corresponding to the range of mid to large {\it positive} hadron rapidities. We have found qualitatively similar results in the three channels. Our results thus provide a conservative uncertainty band for FCEL effects in light hadron nuclear suppression at mid and positive rapidities at LHC.

The $g\to q\bar{q}$ channel can be directly generalized to the production of {\it massive} quarks, $g \to Q\bar{Q}$, relevant to open heavy flavour production in pPb collisions. The effects of FCEL in this process, recently measured at forward rapidity by LHCb~\cite{Aaij:2019lkm,LHCb:2019dpz}, will be computed in a future study.

In order to eventually include FCEL in nPDF global fit analyses, we envisage computing systematically FCEL effects at all rapidities in a perturbative QCD calculation of hadron production in \pA collisions. First, the FCEL calculation presented here should be extended to all possible LO processes (as for instance $g q_{_\textnormal{A}} \to q g$, with $q_{_\textnormal{A}}$ a quark from the target nucleus, 
a channel coming into play at sufficiently backward rapidities). FCEL could then be implemented in a leading-order QCD calculation yielding the relative weights of partonic channels. This program could be also achieved at NLO using FCEL associated to $1\to 3$ forward scattering~\cite{Peigne:2014rka} and NLO calculations of hadron production. 

\acknowledgments
This work is funded by ``Agence Nationale de la Recherche'', grant ANR-COLDLOSS (ANR-18-CE31-0024-02).  

\appendix

\section{Induced coherent radiation spectrum and quenching weight} 
\label{app:quenching} 

The induced coherent radiation spectrum ${\dd I}/{\dd\omega}$ (together with the associated FCEL quenching weight) is a central quantity in our study.
The spectrum associated to $1 \to 1$ forward scattering was previously derived and discussed in Refs.~\cite{Arleo:2010rb,Arleo:2012rs,Peigne:2014uha,Munier:2016oih}. For a generic $1 \to 1$ forward scattering where the incoming parton, outgoing particle, and $t$-channel exchange carry the (Casimir) color charges $C_{\textnormal{in}}$, $C_\R$ and $C_t$ respectively, the coherent radiation spectrum is given by\footnote{The parametric dependence of the spectrum \eq{dIR} was first derived in Ref.~\cite{Arleo:2010rb} for the scattering of a fast color octet undergoing a single hard gluon exchange in the $t$-channel (in which case $F_c = \Nc+\Nc-\Nc = \Nc$). Eq.~\eq{dIR}, together with the rule \eq{color-rule} for the color factor, was shown to hold for any $1 \to 1$ scattering in Ref.~\cite{Peigne:2014uha}.}
\bea
&& \hskip -1cm \omega \frac{\dd I_{_\R}}{\dd \omega}  =  \, F_c \, \frac{\alpha_s}{\pi} \, \left\{ \ln{\left(1+\frac{E^2  \ell_{_{\perp\rm A}}^2}{\omega^2 M_{_\perp}^2}\right)} - \ln{\left(1+\frac{E^2  \ell_{_{\perp \rm p}}^2}{\omega^2 M_{_\perp}^2}\right)} \right\} \, , 
\label{dIR} \\ \nn \\
&& \hskip 2cm F_c = C_{\textnormal{in}} + C_\R - C_t  \, .
\label{color-rule} 
\eea
The {\it induced} spectrum \eq{dIR} is defined for a target nucleus A with respect to a target proton p, and thus vanishes when A$=$p. In \eq{dIR}, $\omega$ and $E$ are the energies of the induced radiation and incoming parton, $M_{_\perp}$ is the transverse mass of the outgoing particle, and $\ell_{_{\perp \rm A}}$ (resp.~$\ell_{_{\perp \rm p}}$) denotes the transverse momentum broadening (to be precisely defined below) across a target nucleus A (resp.~target proton p). 

The associated quenching weight as a function of the energy loss $\varepsilon$ reads~\cite{Arleo:2012rs}
\be
\label{qw-R}
{\cal P}_{_\R}(\varepsilon, E)  = \frac{\dd I_{_\R}}{\dd\varepsilon} \, \exp \left\{ - \int_{\varepsilon}^{\infty} \dd\omega  \frac{\dd{I}_{_\R}}{\dd\omega} \right\} = \frac{\partial}{\partial \varepsilon} \, \exp \left\{ - \int_{\varepsilon}^{\infty} \dd\omega  \frac{\dd{I}_{_\R}}{\dd\omega} \right\} 
\equiv \frac{1}{E} \, \hat{\cal P}_{_\R} \left(\frac{\varepsilon}{E} \right) \, ,
\ee 
where $\hat{\cal P}_{_\R}$ is a scaling function of the ratio 
$x \equiv \varepsilon/E$. Using \eq{dIR} this function can be expressed explicitly in terms of the dilogarithm ${\rm Li}_2(u) = - \int_0^u \frac{\dd v}{v} \ln(1-v)$,
\be
\label{quenching-dilog}
\Phat_{_\R} \left(x, \ell_{_{\perp \rm A}}, M_{_\perp} \right) =  \frac{\partial}{\partial x} \, \exp \left\{F_c \, \frac{\alpha_s}{2 \pi} \left[ {\rm Li}_2\left(\frac{-\ell_{_{\perp \rm A}}^2}{x^2 M_{_\perp}^2} \right) - {\rm Li}_2\left(\frac{-\ell_{_{\perp \rm p}}^2}{x^2 M_{_\perp}^2} \right) \right] \right\}  \, ,
\ee
where the dependence of $\Phat_{_\R}$ on $\ell_{_{\perp \rm A}}$ and $M_{_\perp}$ is now emphasized.  

The nuclear broadening $\ell_{_{\perp \rm A}}$ is the broadening of the radiated gluon across the target (as can be inferred from Ref.~\cite{Peigne:2014uha}). It is related to the average path length $L_{_{\rm A}}$ in the target nucleus,
\be
\label{gluon-broad}
\ell_{_{\perp \rm A}}^2 = \qhat_{_\textnormal{A}} L_{_{\rm A}} \, , 
\ee
where $\qhat_{_\textnormal{A}}$ is the transport coefficient in cold nuclear matter. Being proportional to the gluon distribution in the nuclear target~\cite{Baier:1996sk}, it can be parametrized as~\cite{Arleo:2012rs}
\be
\label{qhat-x}
\hat{q}_{_\textnormal{A}} \equiv \hat{q}_{_0} \left( \frac{10^{-2}}{\tilde{x}_{_2}} \right)^{0.3}\ ;\ \ \ \tilde{x}_{_2} = \min(x_{_0}, \xtwo) \ ; \ \ \  x_{_0} \equiv \frac{1}{2 m_\mathrm{p} L_{_\textnormal{A}}}\, . 
\ee
In \eq{qhat-x}, the parameter $\hat{q}_{_0} \equiv \hat{q}_{_\textnormal{A}}(\tilde{x}_{_2}=10^{-2})$ is already known from $\jpsi$ nuclear suppression data at fixed-target collision energies (as recalled in section~\ref{sec:procedure}), and $\xtwo$ is the longitudinal momentum fraction of the target gluon (normalized by the nucleus mass number $A$) participating to the partonic subprocess as viewed in the \pN\  c.m.~frame. The variable $\tilde{x}_{_2} = \min(x_{_0}, \xtwo)$ occurs because $\hat{q}_{_\textnormal{A}}$ should be evaluated at $\tilde{x}_{_2} \sim \xtwo$ when the hard subprocess is coherent over the nucleus, but at $\tilde{x}_{_2} \sim x_{_0}$ when it is incoherent, and those regimes correspond respectively to $\xtwo < x_{_0}$ and $x_{_0} < \xtwo$~\cite{Arleo:2012rs}.\footnote{In the present study where LHC energies and moderate values of $\pt$ are considered, $\xtwo$ turns out to be always smaller than $x_{_0}$. Thus, in the present study $\hat{q}_{_\textnormal{A}}$ does not depend on $L_{_\textnormal{A}}$ and is a function of $\xtwo$ only, $\hat{q}_{_\textnormal{A}}(\tilde{x}_2) \to \hat{q}(\xtwo)$. We however quote the more general parametrization \eq{qhat-x}, which holds at all collision energies (including fixed-target energies), see Ref.~\cite{Arleo:2012rs}.}

\section{Color state probabilities $\bm{\rho_{_{\rm R}}(\xi)}$}
\label{app:color-proba}

In this Appendix we derive the probabilities $\rho_{_\R}(\xi)$ for the parton pair produced in a $1 \to 2$ forward scattering (as viewed in the target nucleus rest frame) to be in color state $\R$ when the internal energy fractions or partons 1 and 2 are $\xi$ and $1-\xi$. We will consider $g \to gg$, $q \to qg$ and $g \to q \bar{q}$ forward scatterings. When viewed in the proton-nucleon c.m.~frame, those correspond respectively to the $2 \to 2$ partonic processes $gg_{_\textnormal{A}} \to gg$, $qg_{_\textnormal{A}} \to qg$ and $gg_{_\textnormal{A}} \to q \bar{q}$, with $g_{_\textnormal{A}}$ denoting the incoming gluon from the target nucleus. The probabilities are calculated in the general case of the ${\rm SU}(\Nc)$ color group with $\Nc \geq 3$. 

\subsection{\texorpdfstring{${g \to gg}$}{}}
\label{app-g2gg-case}

The $g \to gg$ forward scattering amplitude can be derived most conveniently using light-cone perturbation theory~\cite{Lepage:1980fj} in light-cone $A^+ =0$ gauge. It reads (see also \cite{Gunion:1981qs} and the discussion in \cite{Peigne:2014rka}) 
\be
\label{GBgluon}
{\cal M}_{\rm hard} \  \propto  \ \frac{\Kvec}{\Kvec^{2}} \, \GBugg(14,-45)  + \frac{\Kvec - \qvec}{(\Kvec - \qvec)^{2}} \, \GBtgg(14,-45) - \frac{\Kvec - \xi \qvec}{(\Kvec - \xi \qvec)^{2}} \, \GBsgg(14,-41)  \ \, ,
\ee
where $\qvec$ is the transverse momentum brought by the gluon $g_{_\textnormal{A}}$ from the target nucleus. The graphs in \eq{GBgluon} represent the color factors associated to each Feynman diagram contributing to the amplitude.\footnote{For the pictorial representation of color factors, see for instance Refs.~\cite{Dokshitzer:1995fv,Keppeler:2017kwt}.} From left to right, the graphs in \eq{GBgluon} thus read $T^aT^b$, $\com{T^b}{T^a}$ and $T^bT^a$, where $a$ and $b$ are respectively the color indices of $g_{_\textnormal{A}}$ and of the final gluon 1 (of transverse momentum $\Kvec$ and energy fraction $\xi$, see Fig.~\ref{fig:typical-process}), and $T^a$ the color generators of the ${\rm SU}(\Nc)$ adjoint representation. Note that specifying the overall factor in \eq{GBgluon} is irrelevant for our purpose, since this factor drops out in \eq{proba-def}. 

Using color conservation,\footnote{Note that \eq{color-cons} is often referred to as the Jacobi identity.}
\be
\label{color-cons}
\GBtgg(14,-45) =  \GBsgg(14,-41) -  \GBugg(14,-45)  \ \, , 
\ee
the expression \eq{GBgluon} can be rewritten as
\be
\label{GB-g2gg}
{\cal M}_{\rm hard} \  \propto  \ \left[ \frac{\Kvec - \qvec}{(\Kvec - \qvec)^{2}} - \frac{\Kvec - \xi \qvec}{(\Kvec - \xi \qvec)^{2}} \right] \GBsgg(14,-41) +  \left[ \frac{\Kvec}{\Kvec^{2}} - \frac{\Kvec - \qvec}{(\Kvec - \qvec)^{2}} \right]  \GBugg(14,-45) \, . 
\ee
The probability $\rho_{_\R}$ for the produced gluon pair to be in color state $\R$ is defined by
\be
\label{proba-def}
\rho_{_{\R}} = \frac{|{\cal M}_{\rm hard} \cdot \mathds{P}_{_\R}|^2}{|{\cal M}_{\rm hard}|^2} \, ,
\ee
where $\mathds{P}_{_\R}$ is the hermitian projector on the ($s$-channel) color state $\R$.

We first evaluate the denominator of \eq{proba-def}. By squaring \eq{GB-g2gg} and summing over initial and final color indices we obtain
\be
\label{Mhard-squared}
|{\cal M}_{\rm hard}|^2 = C_{\textnormal{in}} K_{\textnormal{in}} \left\{ \left( C_{\textnormal{in}} -\frac{\Nc}{2} \right) \frac{\xi^2 \qvec^2}{\Kvec^{2}(\Kvec - \xi \qvec)^{2}} +\frac{\Nc}{2}  \frac{\qvec^2}{(\Kvec - \qvec)^{2}} \left[ \frac{1}{\Kvec^{2}} + \frac{(1-\xi)^2}{(\Kvec - \xi \qvec)^{2}}\right]\right\} \, , 
\ee
where $C_{\textnormal{in}}$ and $K_{\textnormal{in}}$ denote the Casimir and dimension of the incoming parton color state, namely, $C_{\textnormal{in}} = \Nc$ and $K_{\textnormal{in}} = \Nc^2-1$ in the present case. 

In order to calculate the numerator of \eq{proba-def}, which depends on the $gg$ color state $\R$, we recall that for $\Nc > 3$, a $gg$ pair can be in six color representations (see \eg~\cite{Dokshitzer:2005ig}), 
\be
\label{glu-glu}
{\bf 8 \otimes 8 = 8_a \oplus 10 \oplus 1 \oplus 8_s \oplus 27 \oplus 0} \, ,
\ee
where `${\bf 10}$' stands for ${\bf 10 \oplus \overline{10}}$, and the representations ordered according to their symmetry properties (${\bf  8_a}$ and ${\bf  10}$ are antisymmetric while ${\bf  1}$, ${\bf  8_s}$, ${\bf  27}$ and ${\bf  0}$ are symmetric) are labelled according to their dimensions when $\Nc =3$. (In particular, ${\bf 0}$ is a symmetric representation which is absent when $\Nc =3$.) For $\Nc >3$ the six color representations $\alpha$ ordered as in the r.h.s.~of \eq{glu-glu} have the following dimensions and Casimirs,
\bea
K_{\alpha} &=& \{ \mbox{\fontsize{12}{2}\selectfont ${\Nc^2-1, \frac{(\Nc^2-1)(\Nc^2-4)}{2}, 1, \Nc^2-1,  \frac{\Nc^2(\Nc-1)(\Nc+3)}{4}, \frac{\Nc^2(\Nc+1)(\Nc-3)}{4}} $} \} \, ,  \\ 
C_{\alpha} &=& \{\Nc,\>2\Nc,\>0,\>\Nc,\>2(\Nc\!+\!1), \>2(\Nc\!-\!1)\} \, .  
\eea

The two color graphs appearing in \eq{GB-g2gg} project on the antisymmetric octet, but respectively in the $s$-channel and $u$-channel of the $2 \to 2$ process, 
\be
\GBsgg(14,-41)  =  \Nc \, \mathds{P}_{\bf  8_a} \ ; \ \ 
\GBugg(14,-45)  = - \Nc \, \mathds{P}_{\bf  8_a}^{(u)} = \frac{\Nc}{2} \mathds{P}_{\bf  8_a}  + 0 \, \mathds{P}_{\bf  10} - \Nc \, \mathds{P}_{\bf  1}  - \frac{\Nc}{2} \mathds{P}_{\bf  8_s} + \mathds{P}_{\bf  27} - \mathds{P}_{\bf  0} \, , \label{gg-8a}
\ee
where the $u$-channel projector $\mathds{P}_{\bf  8_a}^{(u)}$ is written in terms of $s$-channel projectors $\mathds{P}_\alpha$~\cite{Dokshitzer:2005ig}. Inserting \eq{gg-8a} in \eq{GB-g2gg}, the numerator of \eq{proba-def} (summed over initial and final color indices) is obtained using $\mathds{P}_{\alpha} \cdot \mathds{P}_{\beta}^{\dagger} = \mathds{P}_{\alpha} \cdot \mathds{P}_{\beta} = \delta_{\alpha \beta} \, \mathds{P}_{\alpha}$ and $\tr{\mathds{P}_{\alpha}} = K_{\alpha}$. Dividing by \eq{Mhard-squared}, and taking finally the limit $|\qvec| \ll |\Kvec|$ considered in the present study (see section~\ref{sec:setup}), we find the probabilities
\bea
\label{color-proba-gg}
&& \rho_{_{\bf  8_a}} = \frac{\xi^2 + (1-\xi)^2 -1/2}{1+ \xi^2 + (1-\xi)^2} \ \ ; \ \  \rho_{_{\bf 10}} = 0 \ \ ; \ \  \rho_{_{\bf  8_s}} = \frac{1/2}{1+ \xi^2 + (1-\xi)^2}  \ \ ; \nn \\
&& \hskip 10mm \rho_{_{\bf  1}} = \frac{4}{\Nc^2-1} \,  \rho_{_{\bf  8_s}} \ \ ; \ \ \rho_{_{\bf  27}} = \frac{\Nc+3}{\Nc+1} \, \rho_{_{\bf  8_s}} \ \ ; \ \ \rho_{_{\bf  0}} = \frac{\Nc-3}{\Nc-1} \, \rho_{_{\bf  8_s}}  \ \, .
\eea

\begin{figure}[t]
\centering
\includegraphics[width=4.9cm]{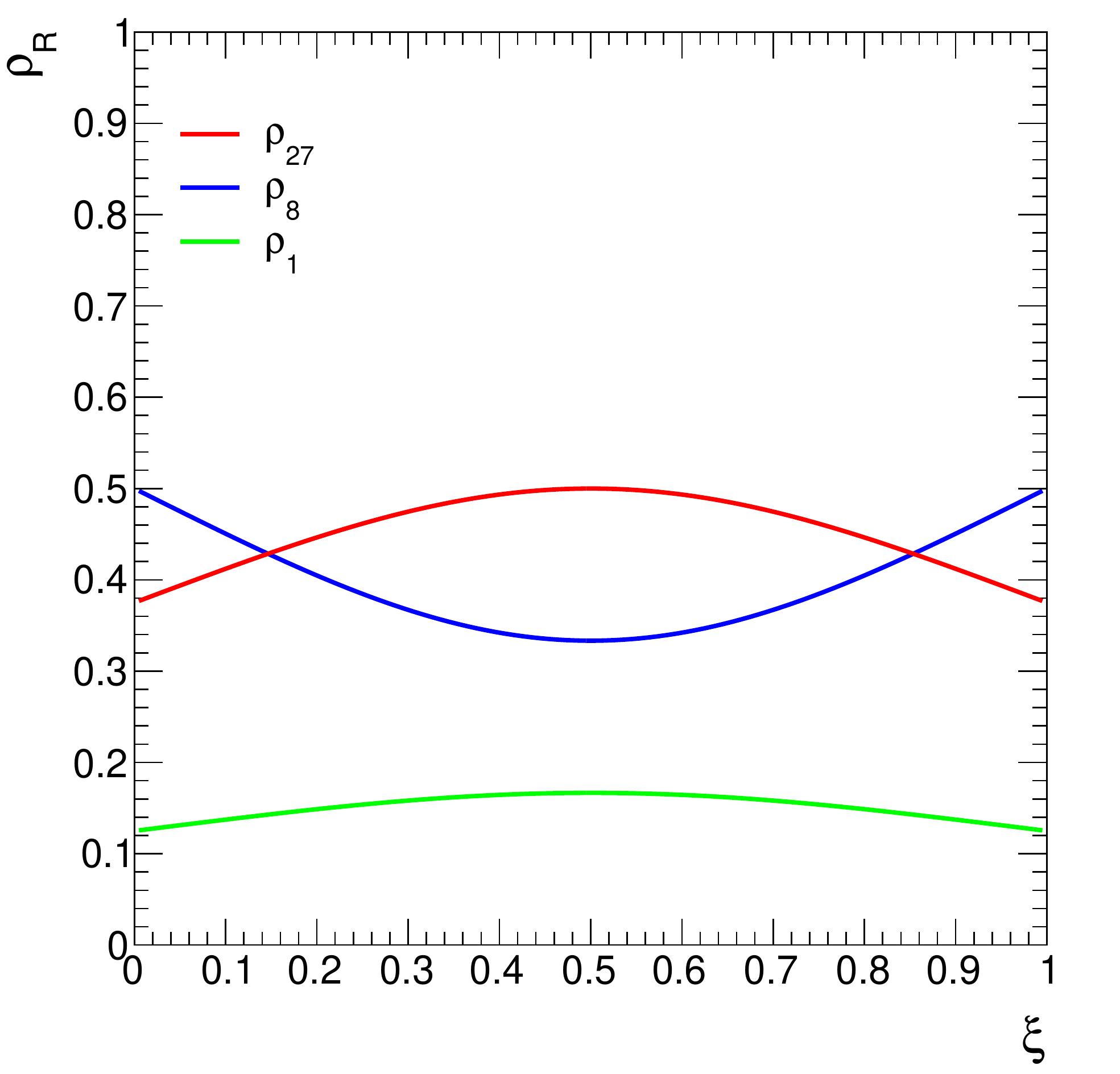} \includegraphics[width=4.9cm]{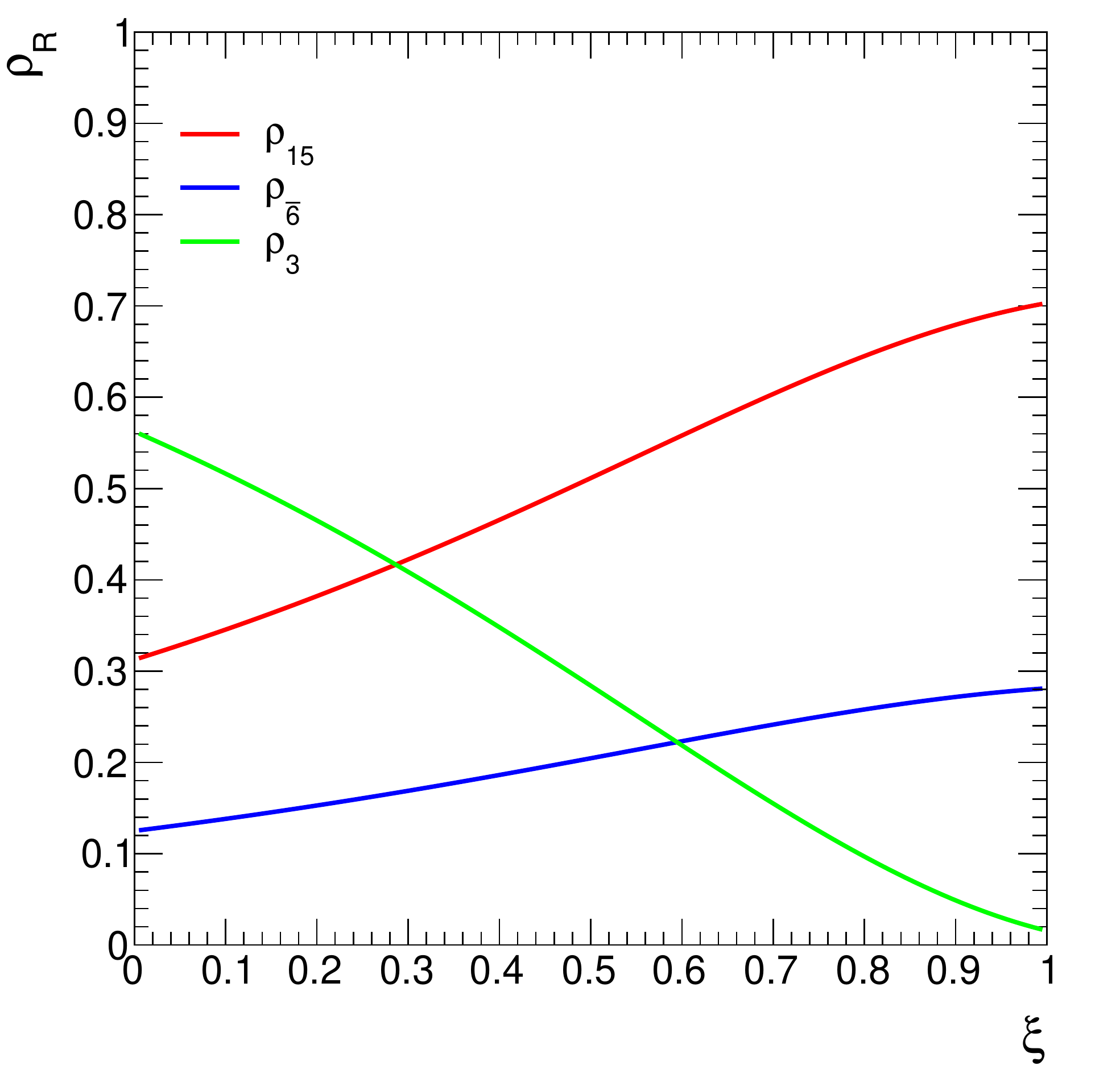} \includegraphics[width=4.9cm]{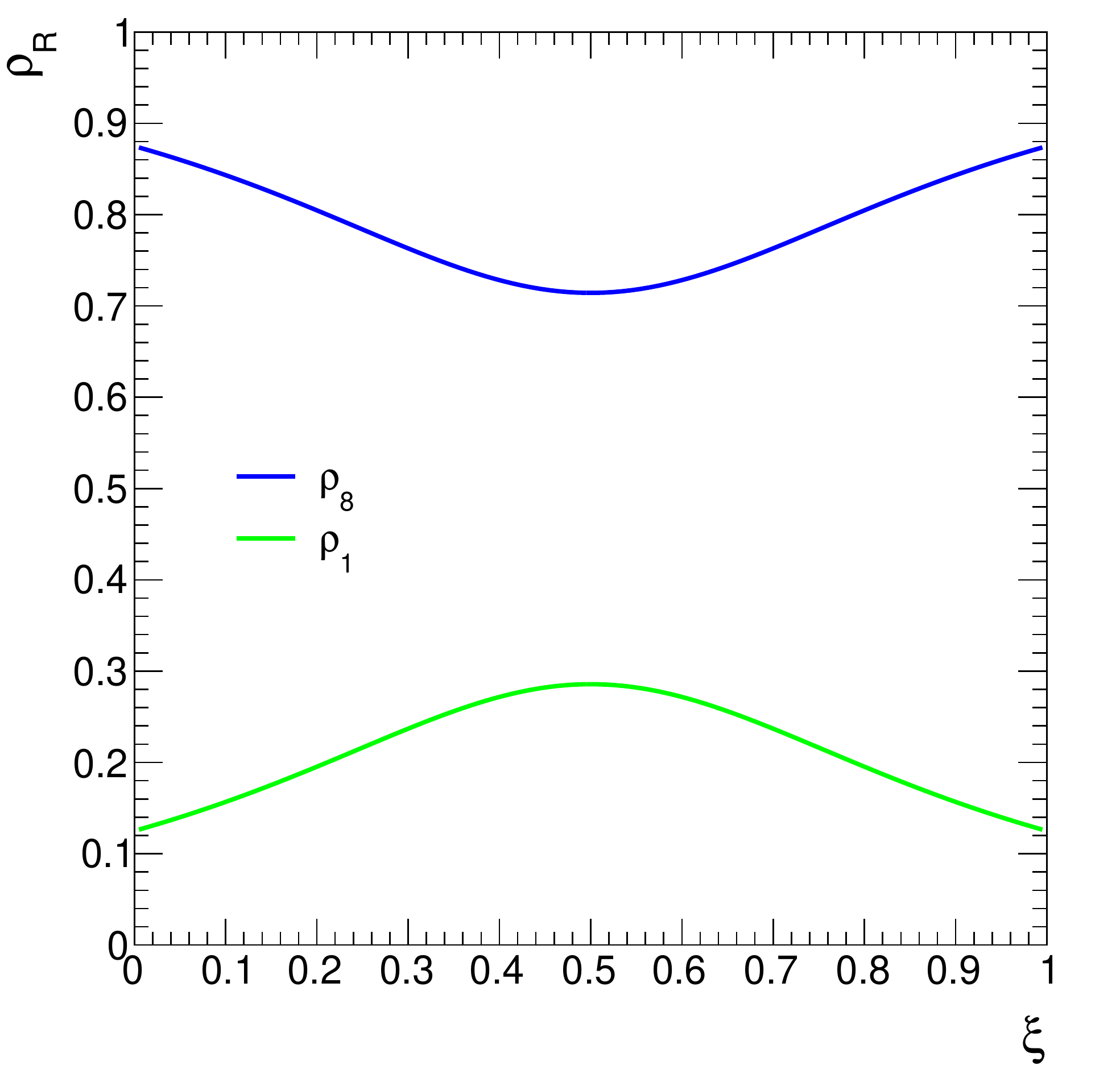}
\caption{Probabilities $\rho_{_{\R}}(\xi)$ for the parton pair produced in $g \to g g$ (left), $q \to q g$ (middle), and $g \to q \bar{q}$ (right) to be in color state $\R$, as a function of the relative energy fraction $\xi$ carried by parton 1 of the pair (chosen to be the gluon in the $q \to q g$ case).} 
\label{fig-probas}
\end{figure}

For $\Nc=3$, the non-vanishing probabilities are given by   
\be
\label{color-proba-pheno-gg}
\rho_{_{\bf 27}}(\xi) = \frac{3/4}{1+ \xi^2 + (1-\xi)^2} \ ;  \ \ \rho_{_{\bf 1}}(\xi) = \frac{1}{3} \rho_{_{\bf 27}}(\xi) \ ; \ \ \rho_{_{\bf  8}}(\xi) = 1 -  \frac{4}{3} \rho_{_{\bf 27}}(\xi)  \ \, ,
\ee
where we have combined the octet representations ${\bf  8_a}$ and ${\bf  8_s}$ which have the same dimension and Casimir. The probabilities \eq{color-proba-pheno-gg} are shown in Fig.~\ref{fig-probas} (left).

\subsection{\texorpdfstring{${q \to qg}$}{}}
\label{app-q2qg-case}

The probabilities $\rho_{_\R}(\xi)$ associated to $q \to qg$ can be derived analogously to the $g \to gg$ case considered previously. We recall that $\xi$ and $\Kvec$ denote the energy fraction and transverse momentum of parton 1 of the produced pair, which for $q \to qg$ is chosen to be the gluon. 

The relations analogous to \eq{GBgluon} and \eq{color-cons} relevant to $q \to qg$ are obtained by replacing the energetic gluon line by a quark line, leading to the analog of \eq{GB-g2gg},
\be
\label{GB-q2qg}
{\cal M}_{\rm hard} \  \propto  \ \left[ \frac{\Kvec - \qvec}{(\Kvec - \qvec)^{2}} - \frac{\Kvec - \xi \qvec}{(\Kvec - \xi \qvec)^{2}} \right]  \GBsqg(16,-45)  +  \left[ \frac{\Kvec}{\Kvec^{2}} - \frac{\Kvec - \qvec}{(\Kvec - \qvec)^{2}} \right] \GBuqg(16,-47)  \, ,
\ee
where the graphs now involve the generators of the ${\rm SU}(\Nc)$ fundamental representation.

It is easy to check that for $q \to qg$ the denominator of \eq{proba-def} is still given by \eq{Mhard-squared}, but with now $C_{\textnormal{in}}= C_F = \frac{\Nc^2-1}{2\Nc}$ and $K_{\textnormal{in}} = \Nc$. For $\Nc \geq 3$, the $qg$ pair can be in three different color states (labelled by their dimensions when $\Nc =3$), 
\be
\label{quark-glu}
{\bf 3 \otimes 8 = 3 \oplus \bar{6} \oplus 15} \, ,
\ee
of dimensions and Casimirs 
\bea
K_{\alpha} &=& \{ \mbox{\fontsize{12}{2}\selectfont ${\Nc, \frac{\Nc (\Nc -2) (\Nc+1)}{2}, \frac{\Nc (\Nc +2) (\Nc-1)}{2}} $} \} \, ,  \\ 
C_{\alpha} &=&  \{ \mbox{\fontsize{12}{2}\selectfont ${C_F,  \frac{(\Nc -1)(3 \Nc+1)}{2\Nc}, \frac{(\Nc +1)(3 \Nc-1)}{2\Nc}} $} \} \, ,
\eea
and associated projectors (satisfying the completeness relation $\sum_\R \mathds{P}_{_\R} = \unit \equiv \! \!$ \QGstraightidentity(12,-20) \ ) 
\bea
\mathds{P}_{\bf 3} &=& \frac{1}{C_F} \, \GBsqg(14,-45)  \ \ \, , \nn \\
\mathds{P}_{\bf \bar{6}} &=&  \frac{1}{2} \, \QGstraightidentity(14,-25)  - \frac{1}{\Nc-1} \, \GBsqg(14,-45)  - \, \GBuqg(14,-47)  \ \ \, ,  \label{quark-glu-proj} \\
\mathds{P}_{\bf 15} &=&  \frac{1}{2} \, \QGstraightidentity(14,-25)  - \frac{1}{\Nc+1} \, \GBsqg(14,-45)  + \, \GBuqg(14,-47)  \ \ \, . \nn
\eea
Using \eq{quark-glu-proj}, each color graph of \eq{GB-q2qg} can be expressed as a linear combination of projectors, namely,
\be
\GBsqg(14,-45) = C_F \mathds{P}_{\bf 3} \ \ ; \ \ \ \ 
\GBuqg(14,-47) = -\frac{1}{2\Nc} \mathds{P}_{\bf 3}  + \frac{1}{2} \left( \mathds{P}_{\bf 15} - \mathds{P}_{\bf \bar{6}} \right)  \, .
\ee
Inserting the latter in \eq{GB-q2qg}, the numerator of \eq{proba-def} is obtained as in the previous section by using $\mathds{P}_{\alpha} \cdot \mathds{P}_{\beta} = \delta_{\alpha \beta} \, \mathds{P}_{\alpha}$ and $\tr{\mathds{P}_{\alpha}} = K_{\alpha}$. After taking the limit $|\qvec| \ll |\Kvec|$, the probabilities associated to $q \to qg$ read
\be
\label{color-proba-qg}
\rho_{_{\bf 15}} = \frac{\frac{\Nc (\Nc +2)}{4(\Nc+1)}}{C_F \xi^2 + \Nc(1-\xi)}  \ \ ; \ \ \rho_{_{\bf \bar{6}}} = \frac{\frac{\Nc (\Nc -2)}{4(\Nc-1)}}{C_F \xi^2 + \Nc(1-\xi)} \ \ ; \ \ \rho_{_{\bf 3}} = \frac{C_F \, (\xi -\frac{\Nc}{2C_F})^2}{C_F \xi^2 + \Nc(1-\xi)} \, .
\ee

For $\Nc=3$, \eq{color-proba-qg} becomes
\be
\label{color-proba-pheno-qg}
\rho_{_{\bf 15}}(\xi) = \frac{15/16}{\frac{4}{3} \xi^2 + 3(1-\xi)} \ ; \ \  \rho_{_{\bf \bar{6}}}(\xi)  = \frac{2}{5} \rho_{_{\bf 15}}(\xi) \ ; \ \ 
\rho_{_{\bf 3}}(\xi) = 1 - \frac{7}{5} \rho_{_{\bf 15}}(\xi)  \, .
\ee
Those probabilities are shown in Fig.~\ref{fig-probas} (middle).

\subsection{\texorpdfstring{${g \to q \bar{q}}$}{}}
\label{app-g2qqbar-case}

As for $q \to qg$, the calculation of the $g \to q \bar{q}$ amplitude in light-cone perturbation theory~\cite{Lepage:1980fj} leads to a result similar to \eq{GBgluon}, up to the appropriate replacement of the color graphs. After using color conservation, we obtain the analog of \eq{GB-g2gg} for the $g \to q \bar{q}$ case:
\be
\label{GB-g2qqbar}
{\cal M}_{\rm hard} \  \propto  \ \left[ \frac{\Kvec - \qvec}{(\Kvec - \qvec)^{2}} - \frac{\Kvec - \xi \qvec}{(\Kvec - \xi \qvec)^{2}} \right]   \GBsqqbar(16,-45) +  \left[ \frac{\Kvec}{\Kvec^{2}} - \frac{\Kvec - \qvec}{(\Kvec - \qvec)^{2}} \right]  \GBuqqbar(16,-47)  \, .
\ee

The final $q \bar{q}$ pair can be projected out on either a singlet or an octet (${\bf 3 \otimes \bar{3} = 1 \oplus 8}$) using the color projectors 
\be
\mathds{P}_{\bf 1}^{q \bar{q}} = \frac{1}{\Nc} \, \FierzIdentity(10,-20)  \ \ ; \ \ \ \ \mathds{P}_{\bf 8}^{q \bar{q}} =  \FierzSinglet(9,-19) - \frac{1}{\Nc} \, \FierzIdentity(10,-20)   \ \, . 
\ee

Using the birdtrack pictorial technique~\cite{Dokshitzer:1995fv,Keppeler:2017kwt}, the calculation of the probabilities \eq{proba-def} associated to $g \to q \bar{q}$ is straightforward and yields (in the limit $|\qvec| \ll |\Kvec|$) 
\be
\label{color-proba-qqbar}
\rho_{_{\bf 1}}^{q \bar{q}}(\xi) = \frac{1}{\Nc^2 (\xi^2 + (1-\xi)^2) -1}  \ \ ; \ \ \rho_{_{\bf 8}}^{q \bar{q}}(\xi) = 1 - \rho_{_{\bf 1}}(\xi) \, .
\ee
The latter probabilities are represented in Fig.~\ref{fig-probas} (right) for $\Nc=3$. 

\section{Parametrization of light hadron cross section in \pp collisions}
\label{app-ppfits}

A main input of the model is the double differential light hadron production cross section in \pp collisions, see~\eq{RpA-yR}. Following the same strategy as in earlier papers on quarkonium production~\cite{Arleo:2012rs,Arleo:2013zua}, the \pp production cross section is fitted by a simple analytic form,
\be
\frac{\dd\sigma_{\pp}^{\psi}}{2 \pi \pt \dd\pt\dd y} \  \propto \  \left(\frac{p_0^2}{p_0^2+\pt^2}\right)^{m} \times \left(1- \frac{2\ \pt}{\sqrt{s}} \cosh{y} \right)^{n} \ .
\label{eq:fit}
\ee
The parametrization is identical to that used in~\cite{Arleo:2013zua}, replacing the quarkonium transverse mass $M_{_\perp}$ in Eq.~(2.11) of~\cite{Arleo:2013zua} by $\pt$ for light hadron production.

\begin{figure}[tbp]
\centering
\includegraphics[width=4.9cm]{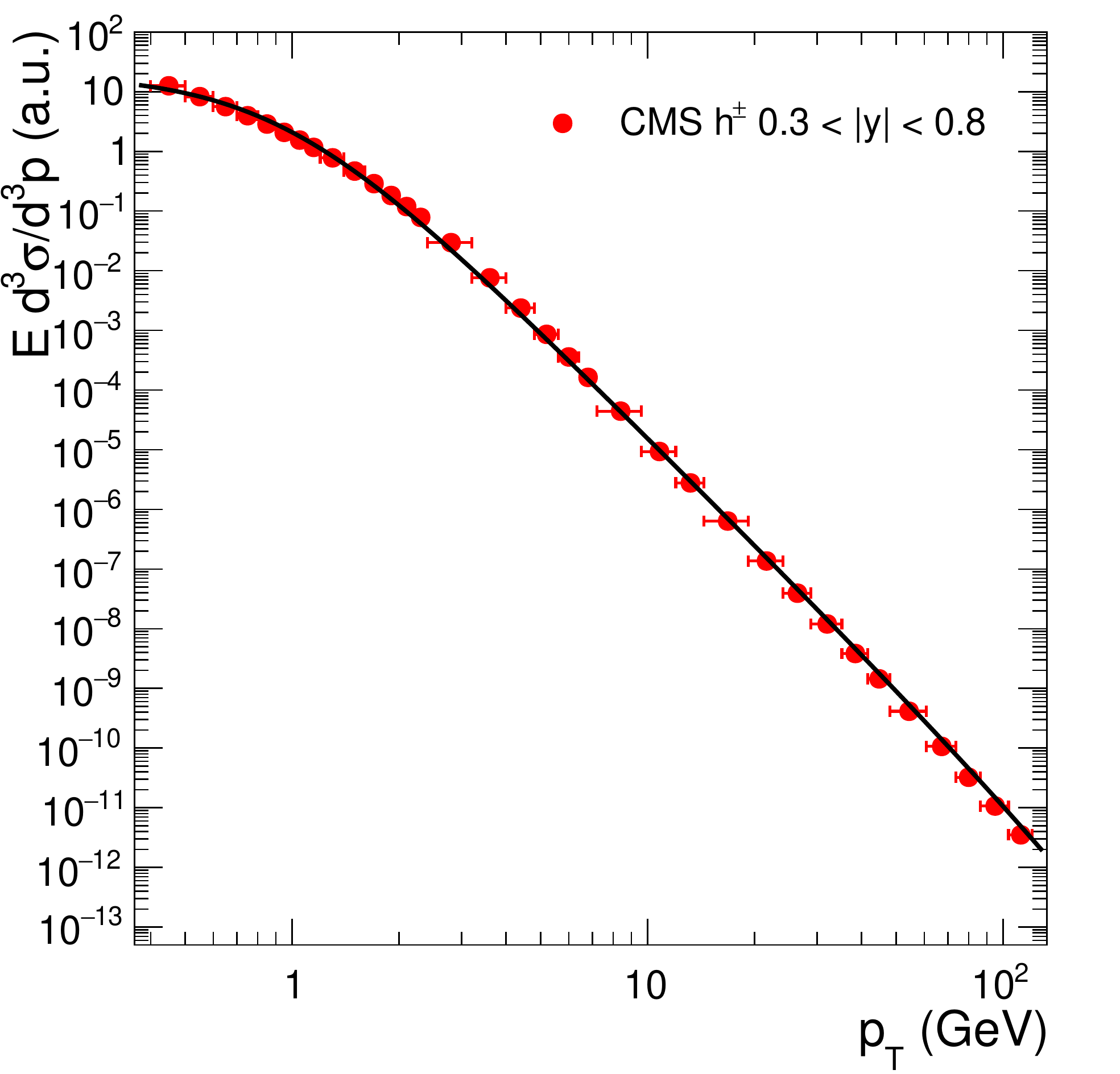}\hskip .2cm
\includegraphics[width=4.9cm]{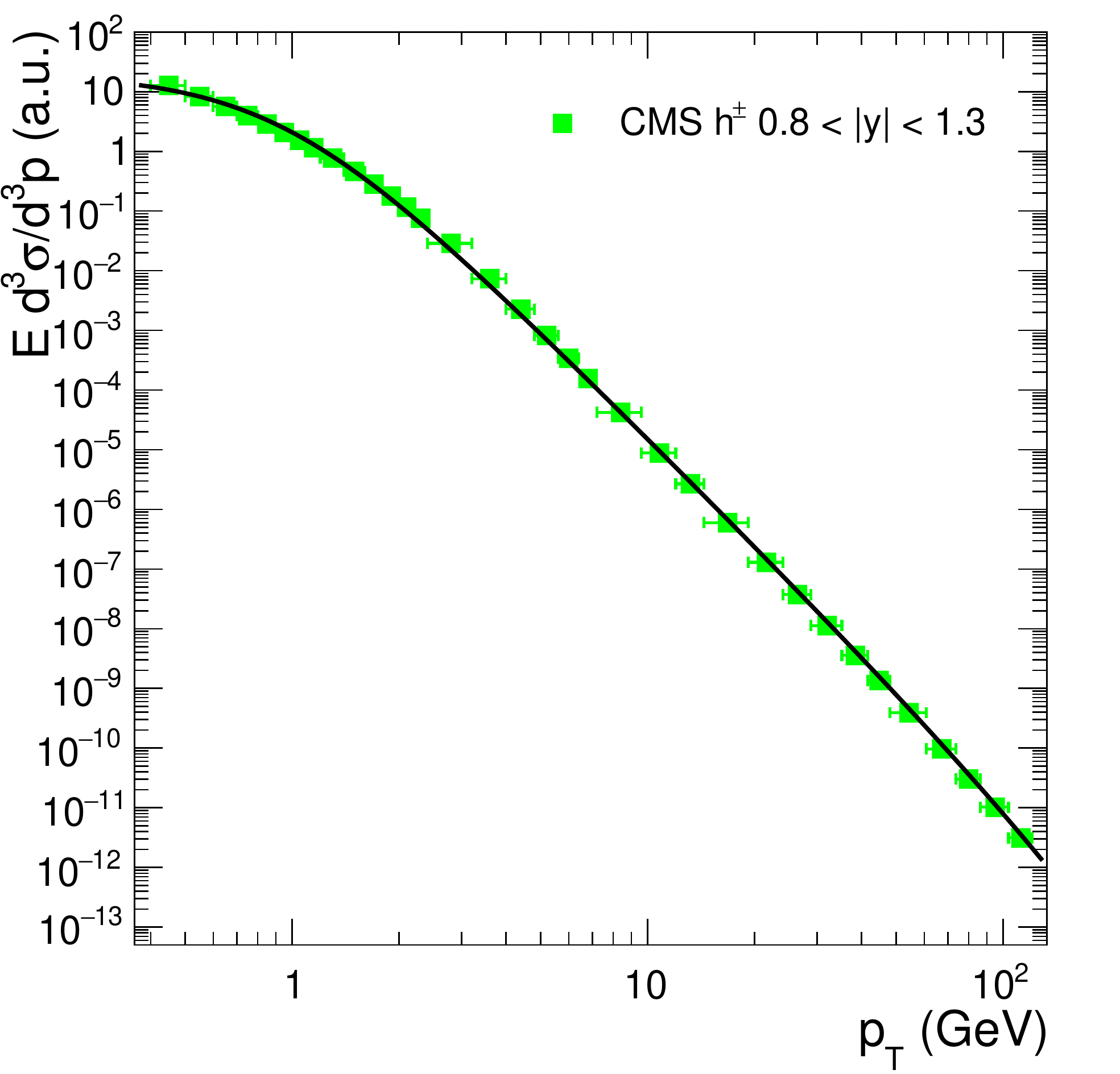}\hskip .2cm
\includegraphics[width=4.9cm]{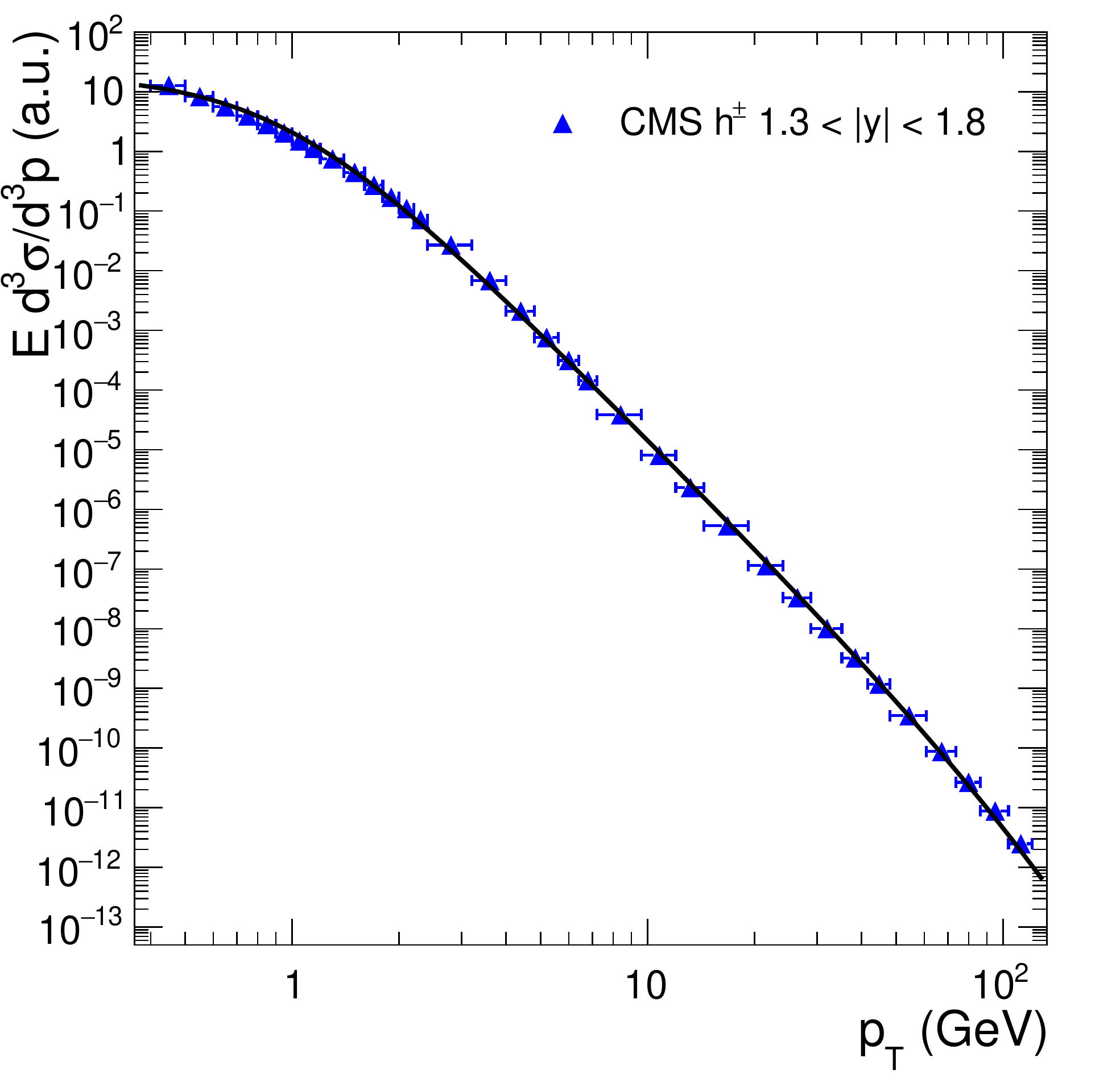}
\caption{Charged hadron spectra measured by CMS in pPb collisions at $\sqrt{s}=5.02$~TeV in the rapidity ranges $0.3 < |y| < 0.8$ (left),  $0.8 < |y| < 1.3$ (center),  $1.3 < |y| < 1.8$ (right)~\cite{Khachatryan:2015xaa}, compared to the parametrization~\eqref{eq:fit}.}
\label{fits-LHC}
\end{figure}

The double differential measurement of the light hadron yields in \pPb\ collisions at $\sqrt{s}=5.02$~TeV has been performed by CMS~\cite{Khachatryan:2015xaa}.\footnote{Other measurements have been performed by ALICE~\cite{Acharya:2018qsh} and ATLAS~\cite{Aad:2016zif} which however lead to looser constraints due to the more restricted $\pt$ range.} The fits to CMS data, shown in Fig.~\ref{fits-LHC} for three intervals in $|y|$, lead to the value $n=15\pm5$. The values of the other parameters ($p_0$ and $m$ in Eq.~\eq{eq:fit}) are irrelevant when computing $R_{\pA}^{h}$, Eq.~\eqref{RpA-y-master}.

The use of pPb instead of pp data in order to parametrize the pp light hadron production rate might seem problematic as these include some nuclear effects. However, those effects have a smooth $y$ and $\pt$ dependence when compared to the absolute pp cross section, resulting in a relatively flat nuclear modification factor (particularly in the $y$ range of the fitted data, $0.3<|y|<1.8$, see for instance Fig.~\ref{fig-RpA-gg}, right) thus affecting the cross section normalization but leaving $n$ unchanged.
We have checked that inferring the pp cross section from the pPb data and such a nuclear modification factor, and using the parametrization~\eq{eq:fit}, provides a value of $n$ which proves fully consistent with the estimate $n=15\pm5$. We also remind that the uncertainty of $R_{\pA}^{h}$ associated to the variation of $n$ is subleading with respect to the other sources of uncertainty 
for rapidities $|y|\lesssim5$, see Fig.~\ref{fig-uncertainties}, right.

\providecommand{\href}[2]{#2}\begingroup\raggedright\endgroup

\end{document}